\documentclass[a4paper,11pt]{article}
\usepackage{jcappub}

\usepackage[export]{adjustbox}
\usepackage[utf8]{inputenc}
\usepackage{color}
\usepackage{graphicx}
\usepackage{amsmath,amsfonts,amssymb}
\hypersetup{breaklinks,colorlinks,urlcolor=blue,citecolor=cyan,linkcolor=magenta}
\usepackage{acronym}
\usepackage{xspace}
\usepackage{multirow}
\usepackage{tabularx}
\usepackage{orcidlink}
\usepackage{ragged2e}
\usepackage{enumitem}
\usepackage{setspace}
\usepackage{booktabs}
\usepackage{comment}
\usepackage{array}

\newcommand{\bea}{\begin{equation}\begin{aligned}} 
\newcommand{\eea}{\end{aligned}\end{equation}}
\newcommand{\be}{\begin{equation}}
\newcommand{\ee}{\end{equation}}

\newcommand{\msun}{M_{\odot}}
\newcommand{\td}{{\rm d}}
\newcommand{\Ndet}{N_{\mathrm{det}}}
\newcommand{\Nobs}{N_{\mathrm{obs}}}
\newcommand{\Rate}{\mathcal{R}}
\newcommand{\mmin}{m_{\min}}
\newcommand{\mmax}{m_{\max}}
\newcommand{\fpbh}{f_{\rm PBH}}

\title{\huge Constraints on primordial black holes \\ from the first part of LIGO-Virgo-KAGRA fourth observing run}
 
\author[a,b,c,\star]{M.~Andr\'es-Carcasona\orcidlink{0000-0002-8738-1672}}
\emailAdd{mandresc@mit.edu}
\author[d]{A.J.~Iovino\orcidlink{0000-0002-8531-5962}}
\author[c]{E.~Vallejo-Pag\`es\orcidlink{0009-0001-8225-5722}}
\author[e,f,g]{V.~Vaskonen\orcidlink{0000-0003-0003-2259}}
\author[e]{H.~Veerm\"ae\orcidlink{0000-0003-1845-1355}}
\author[c,h]{M.~Mart\'inez\orcidlink{0000-0002-3135-945X}}
\author[c]{Ll.~M.~Mir\orcidlink{0000-0002-4276-715X}}
 
\affiliation[a]{LIGO Laboratory, Massachusetts Institute of Technology, Cambridge, MA 02139, USA}
\affiliation[b]{Kavli Institute for Astrophysics and Space Research, Massachusetts Institute of Technology, Cambridge, MA 02139, USA}
\affiliation[c]{Institut de F\'isica d'Altes Energies (IFAE), Barcelona Institute of Science and Technology, E-08193 Barcelona, Spain}
\affiliation[d]{New York University Abu Dhabi, PO Box 129188, Abu Dhabi, UAE}
\affiliation[e]{Keemilise ja Bioloogilise F\"u\"usika Instituut, R\"avala pst. 10, 10143 Tallinn, Estonia}
\affiliation[f]{Dipartimento di Fisica e Astronomia, Universit\`a degli Studi di Padova, Via Marzolo 8, 35131 Padova, Italy}
\affiliation[g]{Istituto Nazionale di Fisica Nucleare, Sezione di Padova, Via Marzolo 8, 35131 Padova, Italy}
\affiliation[h]{Catalan Institution for Research and Advanced Studies (ICREA), E-08010 Barcelona, Spain}
\affiliation[\star]{Corresponding author.}
 
\abstract{We analyse primordial black hole (PBH) populations using state-of-the-art modelling of PBH binaries, deriving the strongest bounds on PBH abundance in the $0.6$--$100\,M_\odot$ range from LIGO-Virgo-KAGRA O4a data and demonstrating sensitivity in the $10^{-4}$--$10^4\,M_\odot$ range, for both monochromatic and log-normal mass functions. The constraints are dominated by resolvable PBH mergers, while the associated gravitational wave background provides complementary but weaker limits. To obtain limits that are agnostic about the astrophysical black hole (ABH) population, we devise two new methods, data-driven methods for statistical inference on scenarios in which PBHs account for a subset of the catalogued events; Allowing for this possibility relaxes slightly the bounds in the solar mass range $2$-$20$ $\msun$. Our bounds are independent of the assumptions about the astrophysical black holes population and represent the most stringent constraints on the PBH abundance in the solar mass range to date.
}

\begin{document}
\maketitle

\section{Introduction} 

The first part of the fourth LIGO-Virgo-Kagra (LVK) observing run (O4a)~\cite{Capote:2024rmo} unveiled 85 new gravitational wave (GW) signals from compact binaries with a false alarm rate of less than one per year~\cite{LIGOScientific:2025slb}. Combined with the results of the first three observation runs (O1-O3)~\cite{LIGOScientific:2018mvr,LIGOScientific:2020ibl,KAGRA:2021vkt}, LVK has detected a total of 161 compact binaries in the O1-O4a integrated observation period. Population studies of this catalogue have revealed interesting features in the black hole (BH) mass distribution, including evidence for multiple peaks above a power-law-like continuum~\cite{LIGOScientific:2025pvj}. 

These findings challenge astrophysical scenarios of BH binary formation. Multiple channels, such as isolated binary evolution and dynamical assembly in dense stellar environments~\cite{Mandel:2018hfr,Woosley:2002zz}, have been proposed, each contributing differently to the astrophysical black hole (ABH) binary population and imprinting distinct mass, spin, and redshift demographics. Accurate modelling of these formation channels requires detailed treatments of stellar evolution, supernova physics, binary interactions, and cluster dynamics, leading to substantial uncertainties in theoretical predictions. Therefore, the current approach in population studies is to parametrise the binary population.

It is possible that some features of the observed binary BH population may not be fully explained by the ABH channels but may instead originate from a population of primordial BHs (PBHs). Such objects, hypothesised to have formed in the early Universe from the collapse of large density fluctuations~\cite{Carr:1974nx}, would constitute a fundamentally different BH population. If present, PBHs could imprint characteristic signatures on the merger rate, mass spectrum, spin distribution, and redshift evolution of binary BHs, potentially accounting for features that are difficult to reconcile with ABH scenarios alone.

Previous analyses of the LVK catalogue have not identified statistically significant evidence for a PBH component in the observed binary population~\cite{Hutsi:2020sol, Hall:2020daa, Wong:2020yig, Franciolini:2021tla, DeLuca:2021wjr, Andres-Carcasona:2024wqk,Afroz:2024fzp}. Instead, the absence of a clear PBH signal has been translated into stringent upper limits on the abundance of PBHs in the solar mass range~\cite{Hutsi:2020sol, Andres-Carcasona:2024wqk}, typically expressed as the fraction of dark matter (DM) they can constitute. 

Since a robust quantitative model of the ABH binary population is currently lacking, we will, however, refrain from looking for specific PBH related signatures and consider ABHs as a potential foreground that affects how LVK observations constrain the PBH population. We propose novel approaches for dealing with ABHs in the most agnostic way. This is needed to obtain the most conservative constraints on PBHs that do not depend on the uncertainties of ABH models.

In our earlier work~\cite{Andres-Carcasona:2024wqk}, we derived constraints on the abundance of PBHs using the LVK catalogue up to the third observing run. Beyond adding the new data from O4a, the present analysis extends the scope of~\cite{Andres-Carcasona:2024wqk} in three important ways: (i) by introducing a new way of estimating the model agnostic constraint, (ii) by incorporating contributions from galactic binaries~\cite{Pujolas:2021yaw}, and (iii) by deriving the constraints implied by the non-observation of a stochastic GW background (SGWB)~\cite{Mukherjee:2021ags,Raidal:2017mfl,Boybeyi:2024mhp,Bavera:2021wmw,Jiang:2024aju,Romero-Rodriguez:2024ldc,Sah:2025agw}.

We use geometric units $G=c=1$ and the Planck 2018 values for cosmological parameters~\cite{Planck:2018vyg}.

\section{Methods}
\label{sec:methods}

The GW signatures of BH binaries stem from their merger rate. For the most common PBH scenarios, the merger rate can be estimated analytically if the PBH mass function is not too wide and the PBHs are not initially formed in clusters.\footnote{These assumptions are relatively mild, as strong initial clustering is difficult to achieve in critical collapse scenarios~\cite{Crescimbeni:2025ywm}, and the PBH constraints tend to prefer narrow mass functions~\cite{Carr:2017jsz,Carr:2026hot}.}
We consider the early two- and three-body channels for PBH binary formation~\cite{Nakamura:1997sm,Ioka:1998nz,Raidal:2017mfl,Ali-Haimoud:2017rtz,Raidal:2018bbj,Vaskonen:2019jpv,Hutsi:2020sol,DeLuca:2020fpg,DeLuca:2021pls,Delos:2024poq,Raidal:2024bmm}, summarised in Appendix~\ref{app:PBHBs}, and omit PBH binary formation in the late universe, which gives a subdominant contribution to the merger rate~\cite{Raidal:2017mfl,DeLuca:2020jug,Raidal:2024bmm}.

The shape of the mass function of the PBH population depends on the model of PBH formation and, e.g., in critical collapse scenarios, can possess additional features due to the softening of the equation of state during the QCD phase transition, which enhances the PBH abundance around a solar mass~\cite{Jedamzik:1996mr, Byrnes:2018clq, Franciolini:2022tfm, Escriva:2022bwe, Musco:2023dak,Ferrante:2022mui}. However, as shown in~\cite{Andres-Carcasona:2024wqk}, the constraints on the fraction of DM in PBHs, $f_{\rm PBH} \equiv \Omega_{\rm PBH}/\Omega_{\rm DM}$, are primarily dependent on the average PBH mass $\langle m \rangle$. To account for variations in the width of the mass function, we consider log-normal mass functions
\be
    \psi \propto \exp\left[ -\frac{\ln^2(m/m_c)}{2\sigma^2} \right]\,,
\ee
where $m_c$ and $\sigma^2$ denote the mode and the width of the distribution. It is a good approximation of wider peaks in the mass spectrum and is predicted by several PBH formation mechanisms~\cite{Carr:2026hot}. We vary the mode $m_c$ across the LVK-sensitive range (sub-solar to stellar masses), where the means of these distributions are given by $\langle m \rangle = m_c e^{-\sigma^2/2}$. The limiting case $\sigma \to 0$ gives the monochromatic mass function, $\psi \propto \delta(m-\langle m \rangle)$.

\subsection{Resolvable binaries}

For resolvable binaries, we adopt a hierarchical Bayesian approach. The associated likelihood is~\cite{Thrane:2018qnx, Mandel:2018mve, LIGOScientific:2020kqk, Mastrogiovanni:2023zbw}
\be \label{eq:hierarchicalBayesian}
    \mathcal{L}(\{d\}|\Lambda) \propto e^{-N(\Lambda)}\prod_{i=1}^{\Nobs}\int \mathcal{L}(d_i|\theta)\pi(\theta|\Lambda)\td \theta\,,
\ee
where $\{d\}$ denotes the data, $\Nobs$ the number of observed events, and $N(\Lambda)$ the expected number of observed events in a BH binary population characterised by the model parameters contained in $\Lambda$, which include, for instance, the PBH abundance and the parameters describing their mass function. The event likelihood $\mathcal{L}(d_i|\theta)$ is given in terms of the event parameters $\theta$: the masses $m_1$ and $m_2$, and the redshift $z$, whose distribution is defined by the hyperprior $\pi(\theta|\Lambda)$. We omit the spins in our analysis since we expect them to have a minor effect on the PBH constraints. The hyperprior is derived from the BH merger rate $\Rate$ as
\be \label{eq:hyperp}
    \pi(\theta|\Lambda) 
    = \frac{1}{1+z}\frac{\td V_c}{\td z} \frac{\td \Rate}{\td m_1 \td m_2}(\theta|\Lambda) \, ,
\ee
where $V_c$ is the comoving volume. The expected number of observable events is
\be\label{eq:Nexpected}
    N(\Lambda) = T\int p_{\mathrm{det}}(\theta)\pi(\theta|\Lambda)\td \theta\, ,
\ee
where $p_{\mathrm{det}}(\theta)$ denotes the detection probability that depends primarily on the masses and redshift of the binary~\cite{Finn:1992xs,Gerosa:2019dbe} and $T$ the observing time. We estimate the detection probability following the signal-to-noise (SNR) based prescription stated in Refs.~\cite{Hutsi:2020sol,Andres-Carcasona:2024wqk} with a threshold ${\rm SNR}_{\rm thr} = 8$ and the representative noise amplitude spectral densities from Ref.~\cite{LIGOScientific:2025hdt}. The numerical implementation of the likelihood~\eqref{eq:hierarchicalBayesian} is described in Appendix~\ref{app:methods}.

Although there has been some speculation about the potential primordial origin of some more extreme GW events~\cite{Yuan:2025avq,DeLuca:2025fln}, an astrophysical origin cannot be ruled out for any of the events detected so far~\cite{Bartos:2025pkv,Croon:2025gol,Kiroglu:2025vqy}. It is generally challenging to confirm an astrophysical origin for individual events. Deciding whether an event stems from astrophysics is important not only for the identification of candidate PBHs but also for imposing reliable constraints on PBHs. 

The constraints on PBHs are inevitably weaker if any of the observed GW events have a primordial origin because this allows for larger merger rates. This is the effect we must incorporate into our analysis -- we aim to estimate the constraints on the PBH abundance without knowing whether some of the observed BH binary merger events are primordial or not. A robust quantitative model for ABH is not required for this. On the contrary, the most conservative constraints on the PBH abundance arise when as little as possible is assumed about the ABH population. In the following, when deriving constraints on the abundance of PBHs, the ABH population is approached in three different ways:
\begin{enumerate}[leftmargin=*]
    \item In the \emph{astrophysics only} case, all the observed events originate from ABH binaries, and no PBH are assumed to have been observed. In this case, the constraint on $f_{\rm PBH}$ is derived from the non-observation of any excess of events. The 95\% CL upper bound on $f_{\rm PBH}$ corresponds to the region where $N(\Lambda) < 3$.
    
    \item In the \emph{astrophysically agnostic} case, any of the observed events could be primordial. We therefore set limits without modelling the ABH population, using two complementary procedures: (i) taking random subsets of the catalogue and assuming that they are PBHs, as done in Refs.~\cite{Hutsi:2020sol,Andres-Carcasona:2024wqk}; and (ii) using a subset–marginalised likelihood (SML), which smoothly down–weights events inconsistent with the PBH population. Implementation details and a derivation of this new procedure are provided in Appendix~\ref{app:agnosticABHs}.
    
    \item In the \emph{astrophysically informed} case, we model all observed events from a merger rate that combines contributions from both PBHs and ABHs. For BH binaries formed through astrophysical channels, a reliable estimate of the merger rate based on first principles is missing, which is why we adopt the phenomenological power-law + 2 peaks (PL+2P) parametrisation~\cite{Mastrogiovanni:2023zbw, LIGOScientific:2025jau,LIGOScientific:2020kqk} to model the ABH population. Details are given in Appendix~\ref{app:parametrizedABHs}.   
\end{enumerate}
\subsection{Subsolar mass binaries}
Although signals from individual extragalactic subsolar-mass binaries are generally too weak to be detectable by the LVK detectors~\cite{LIGOScientific:2019kan,LIGOScientific:2021job,LVK:2022ydq,Nitz:2022ltl,Nitz:2021mzz,Andres-Carcasona:2022prl,LIGOScientific:2026wxz}, the inspiral phase of much closer Galactic binaries could still be observable. To model this possibility, we adapt the approach of Ref.~\cite{Pujolas:2021yaw}, in which the spatial distribution of galactic PBH binaries is modelled using the early two- and three-body merger rate scaled with the DM density of the Milky Way halo. For the latter, we adopt a Navarro–Frenk–White profile with parameter values from~\cite{Cautun:2019eaf}.

Various methods have been developed to target signals from galactic planetary-mass binaries~\cite{Miller:2020kmv,Miller:2021knj,Miller:2024jpo,Andres-Carcasona:2024jvz,Andres-Carcasona:2023zny,Alestas:2024ubs}, and all searches to date have yielded null results~\cite{Miller:2024fpo,LIGOScientific:2025vwc}. The most effective strategy to detect a monochromatic signal of unknown frequency relies on computing the power spectrum and identifying statistically significant peaks above the expected noise background. To estimate the current sensitivity to such light binaries within the Milky Way, we consider the optimal SNR\footnote{Here, “optimal SNR” refers to the fully coherent sensitivity of an ideal periodogram over the full observing time $T_{\rm obs}$ for an effectively monochromatic signal that remains within one Fourier bin (after any demodulation). In practice, real searches employ semi-coherent methods and other data-processing techniques that reduce the effective sensitivity below this optimal value.}, obtained from a periodogram of length $T_{\mathrm{obs}}$ given by~\cite{Astone:2014esa}
\be \label{eq:SNRopt_cw}
    {\rm SNR}_{\rm opt} = \frac{h_0}{2} \sqrt{\frac{T_{\rm obs}}{S_n(f_0)}}\,,
\ee
where $h_0$ is the characteristic strain of the monochromatic signal at frequency $f_0$. Using the leading-order post-Newtonian correction, valid in the regime of low velocities and weak gravity, the strain can be approximated as
\be
    h_0 = \frac{4}{D_{\rm L}} \mathcal{M}_c^{5/3} (\pi f_0)^{2/3}\,,   
\ee
where $D_{\rm L}$ denotes the luminosity distance to the binary and $\mathcal{M}_c$ its chirp mass. The signal will have a slowly evolving frequency. Still, many detection methods rely on a heterodyne correction to convert it into a purely monochromatic signal and increase its SNR for enhanced sensitivity~\cite{Andres-Carcasona:2024jvz,Miller:2024jpo,Andres-Carcasona:2023zny}.
The time to coalescence $\tau$ for a binary emitting at frequency $f_0$ is given, in the leading-order quadrupole approximation~\cite{Maggiore:2007ulw}
\be
    \tau = \frac{15}{768} \mathcal{M}_c^{-5/3} (\pi f_0)^{-8/3} \, .
\ee
For $T_{\rm obs}$ in Eq.~\eqref{eq:SNRopt_cw} we use the minimum between the total observing time, $T$, and $\tau$ and we adopt the standard detection threshold $\rm SNR \geq 8$~\cite{Pujolas:2021yaw}. This value can also be related to an equivalent $5\sigma$ detection probability using the peakmap sensitivity estimate derived in Ref.~\cite{Astone:2014esa}. In the fully coherent limit, the optimal SNR is
\be
{\rm SNR}_{\rm opt} = \frac{\mathcal{B}}{2} \sqrt{ \mathrm{CR}_{\rm thr} -\sqrt{2}\operatorname{erfc}^{-1}(2\Gamma) } = \frac{\mathcal{B}}{2} \sqrt{\mathrm{CR}_{\rm thr}+z_\Gamma}, ,
\ee
where $\Gamma$ is the detection probability and
$$
z_\Gamma \equiv -\sqrt{2}\mathrm{erfc}^{-1}(2\Gamma)
$$
is the corresponding one-sided Gaussian quantile. The quantity $\mathrm{CR}_{\rm thr}$ is the critical ratio threshold used to select statistically significant candidates from the frequency projection of the peakmap. The coefficient $\mathcal{B}$ accounts for the threshold used to select individual peaks in the peakmap and for the averaging over the source sky position and polarization. For an all-sky search, $\mathcal{B}\simeq4.97$~\cite{Astone:2014esa}. Taking $\mathrm{CR}{\rm thr}=5$~\cite{Astone:2014esa} and $z_\Gamma=5$, corresponding to a one-sided $5\sigma$ detection probability, gives ${\rm SNR}_{\rm opt} \simeq 7.9$, consistent with the adopted threshold.

\subsection{Stochastic GW background}

Unresolvable binaries contribute to a SGWB, whose current non-detection in the LVK band~\cite{LIGOScientific:2016jlg,LIGOScientific:2019vic,KAGRA:2021kbb,LIGOScientific:2025bgj,KAGRA:2021mth,KAGRA:2021rmt,Romero-Rodriguez:2021aws,LIGOScientific:2025kry} provides another way to constrain the PBH abundance~\cite{Hutsi:2020sol,Wang:2016ana,Atal:2022zux,Bavera:2021wmw,Boybeyi:2024mhp,Mukherjee:2021ags,Raidal:2017mfl,Jiang:2024aju,Romero-Rodriguez:2024ldc,Zhou:2024yke,LIGOScientific:2025kry}. This method is interesting as it is completely independent of the ABH population. The abundance of the SGWB per logarithmic frequency interval is given by
\be \label{eq:Omega_GW}
    \Omega(f) 
    \!=\!\frac{1}{\rho_c}\int \frac{1+z}{4\pi D_{\rm L}^2} 
    \frac{\td E}{\td \ln f_s} \left(1- p_{\rm det}(\theta)\right)\!\pi(\theta|\Lambda)  \td \theta,
\ee
where $\rho_c = 3H_0^2/8\pi$, $H(z)$ is the Hubble expansion rate and $\td E_{\rm GW}/\td \ln f_s$ is the energy spectrum emitted by the binary, evaluated at the source frequency $f_s = (1+z)f$. The factor $1- p_{\rm det}(\theta)$ is included to remove the contribution from the resolvable binaries. Details on the estimate are gathered in Appendix~\ref{app:SGWB}.

For an isotropic background $\Omega(f)$, the optimal cross-correlation statistic for a detector pair $I\!J$ has an expected SNR of~\cite{Christensen:1992wi,Allen:1997ad,Romano:2016dpx}
\be \label{eq:SNR_cc_SGWB}
    {\rm SNR}_{IJ}
    = \frac{3 H_0^2}{10\pi^2}\,
    \left[2T\int_{0}^{\infty}\td f\;
    \frac{\gamma_{IJ}^2(f)\Omega^2(f)}{f^{6}P_I(f)P_J(f)}\right]^{1/2}.
\ee
Here $T$ is the coincident observation time for the baseline $IJ$, $P_I(f)$, and $P_J(f)$ are the noise PSDs of the two detectors, and $\gamma_{IJ}(f)$ is the overlap-reduction function. Eq.~\eqref{eq:SNR_cc_SGWB} assumes stationary, Gaussian, and uncorrelated noise between detectors and that the same segmentation and windowing are used to estimate $P_{I,J}(f)$. Since we combine multiple baselines between the LIGO and Virgo detectors and multiple observing periods with different sensitivities, we add the SNR in quadrature to obtain the network SNR as
$
    {\rm SNR}^2 = \sum_{e,I\neq J}{\rm SNR}_{IJ,e}^2\, ,
$
where the epoch index $e$ carries its own time, $T_e$, and noise spectrum $P_{I,J}^{(e)}(f)$. We consider the SGWB detectable for an $\rm SNR\geq 8$~\cite{Romero-Rodriguez:2024ldc,Pujolas:2021yaw}.

\section{Results}

\begin{figure*}
    \centering
    \vspace{-10pt}
    \includegraphics[width=\textwidth]{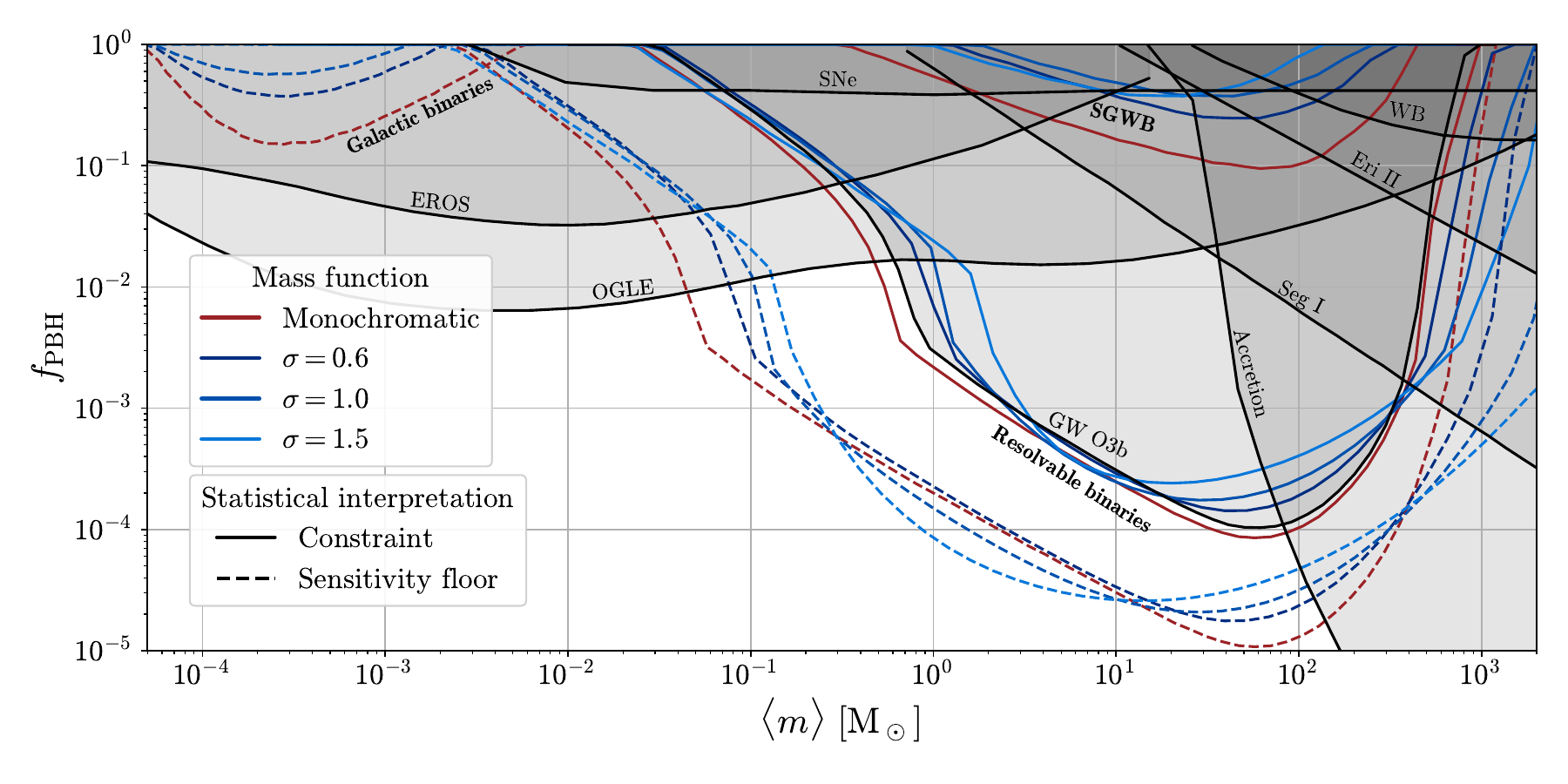}
    \caption{Constraints on the DM fraction of PBHs for monochromatic and log-normal mass functions. The other constraints in gray are computed for monochromatic mass functions. They include GW O3b~\cite{Andres-Carcasona:2024wqk}, EROS~\cite{EROS-2:2006ryy}, OGLE~\cite{Mroz:2024mse,Mroz:2024wag}, Seg1~\cite{Koushiappas:2017chw}, Accretion~\cite{Serpico:2020ehh}, Eri II~\cite{Brandt:2016aco}, WB~\cite{Monroy-Rodriguez:2014ula} and SNe~\cite{Zumalacarregui:2017qqd}. The solid lines show upper limits from resolvable binaries and SGWB, while the dashed lines show the sensitivity floor.}
    \label{fig:LimitsO4a_Combined}
\end{figure*}

The constraints from the non-observation of resolvable PBH binaries in the astrophysics only case and from the non-observation of a SGWB are shown in Fig.~\ref{fig:LimitsO4a_Combined} by the solid lines for the monochromatic and log-normal mass functions. The bounds on $f_{\rm PBH}$ from the non-observation of resolvable PBH binaries exceed those from the non-observation of a SGWB from unresolvable PBH binaries by about three orders of magnitude. The O4a data slightly improve the O3 constraint from resolvable binaries (shown in black for the monochromatic case), providing the tightest constraints in the mass region $0.5-90~\msun$, for which the upper bounds in  $\fpbh$ vary between  $10^{-2}$ and $10^{-4}$, respectively. For broader mass functions, the constraint extends to higher masses and becomes weaker around $\langle m \rangle \sim 100\,\msun$. Qualitatively, this result aligns with the general trend in the PBH constraints for extended mass functions, as reported in~\cite{Carr:2017jsz,Carr:2026hot}. However, the method of~\cite{Carr:2017jsz} cannot be applied directly here, as PBHs of different masses do not contribute independently.

Although our analysis of galactic subsolar mass binaries does not yield a new constraint on the PBH abundance, it allows us to identify a sensitivity region where a detection would have been expected given the current LVK sensitivity.
Indeed, under our assumptions, observing PBH binaries would be consistent at 95\% CL with PBH models when $f_{\rm PBH} \gtrsim 0.1$ around the mass range $\langle m\rangle \in (10^{-4}\msun,10^{-3}\msun)$. This sensitivity floor is shown by the dashed lines in Fig.~\ref{fig:LimitsO4a_Combined}, and the region above it corresponds to $f_{\rm PBH}$ values for which the expected number of observed events is $N(\Lambda) > 0.05$. For galactic binaries, this region of the parameter space is, however, excluded by EROS and OGLE microlensing bounds.

\begin{figure}
    \centering
    \vspace{-10pt}
    \includegraphics[width=0.6\columnwidth]{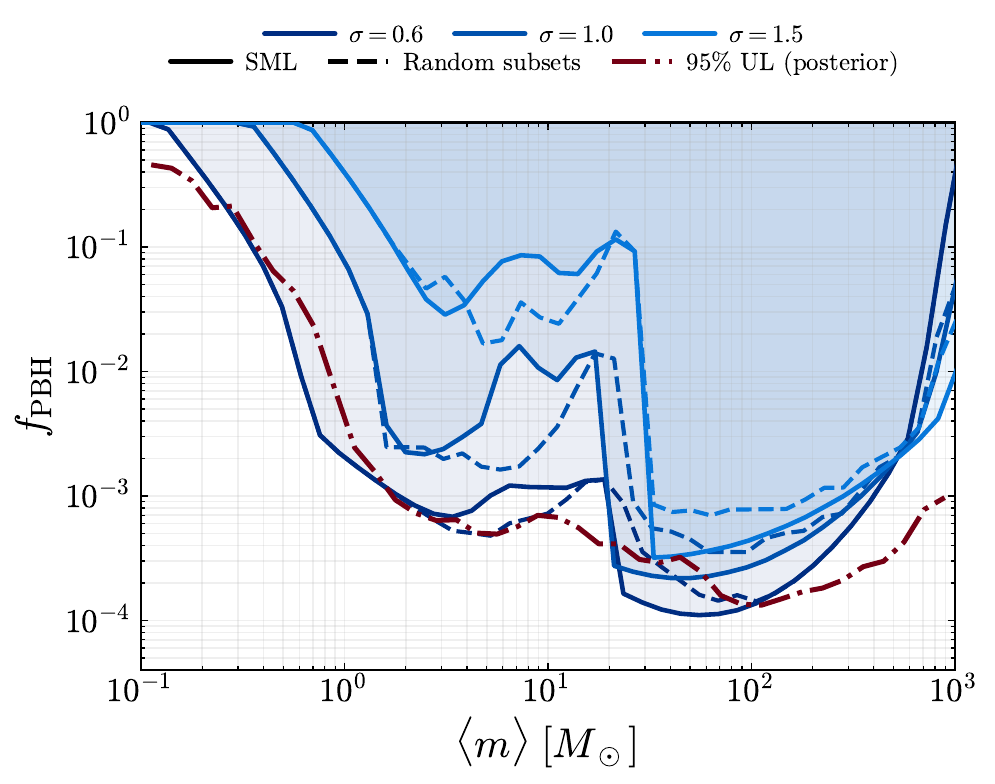}
    \caption{Comparison of the agnostic limits obtained with both methods and the 95\% upper limits obtained from the posterior samples of the PBH+ABH fit assuming the PL+2P and log-normal with $\sigma=0.6$. The minimal ABH mass is fixed to $m_{\min}=5~M_\odot$ and the event GW200210\_092254 is omitted.}
    \label{fig:withABH}
\end{figure}

Allowing for the possibility that a subset of the catalogue is primordial, Fig.~\ref{fig:withABH} compares two agnostic procedures. The procedure based on random subsets, constructed with $10^5$ combinations, yields conservative limits by construction, and the constraint converges as more combinations are sampled. On the other hand, the subset–marginalised likelihood produces comparable constraints while smoothly down–weighting events that are disfavoured by the PBH model (for details, see Appendix~\ref{app:agnosticABHs}). Both methods show that when some of the events are allowed to be primordial, the constraints relax in the region $2~\msun < \langle m\rangle < 20~\msun$. Increasing the width $\sigma$ does not qualitatively alter the behaviour, but it relaxes the constraints on the PBH abundance. For a $\sigma = 1.5$, these lower to $\fpbh\sim 10^{-1}$, while for $\sigma = 1.0$ they reduce to $\fpbh\sim 10^{-2}$, and for $\sigma=0.6$ to $\fpbh\sim 10^{-3}$. Since these agnostic constraints themselves do not rely on any ABH rate model, they provide robust, model-independent bounds on $f_{\rm PBH}$. While wider mass functions seem to allow for a larger fraction of PBHs, these models are ruled out by other experiments, such as CMB constraints~\cite{Serpico:2020ehh,Poulin:2017bwe}.

When we model the ABH population with the PL+2P parametrisation and add a PBH component with a log–normal mass function with $\sigma=0.6$, the Bayesian fit tends to assign the lowest mass events to the PBH population and utilise the ABH population to account for the more massive events, with $m \gtrsim  7~\msun$ (for details, see Appendix~\ref{app:parametrizedABHs}). This yields a log-Bayes factor of $\ln \mathcal{B}\simeq 30.5$, indicating very strong evidence towards the PBH+ABH hypothesis under the models being employed. Upon closer inspection of the individual likelihoods of the events (listed in Appendix~\ref{app:parametrizedABHs}), only four events are actually selected to be primordial. These are events that have at least one very light component, below $\approx 7~\msun$.
To avoid overinterpreting this phenomenological flexibility, we fix $m_{\min}=5\,M_\odot$ and omit GW200210\_092254, whose lighter component might be a neutron star, in our fiducial inference. By only changing this, the Bayes factor no longer provides evidence for a PBH contribution, yielding a log-Bayes factor of $\ln \mathcal{B} \simeq 1.7$. From the posterior samples of this fit, we obtain a 95\% upper limit on $\fpbh$ as a function of $\langle m \rangle$. As seen from Fig.~\ref{fig:withABH}, this constraint is qualitatively similar to the purely agnostic ones.

Moreover, the mixed ABH+PBH inferences are highly sensitive to seemingly modest analysis choices. This behaviour reflects the limited identifiability between a flexible parametrisation of the ABH population and a subdominant PBH population. Consequently, apparent evidence for PBHs can be prior- or selection-driven rather than data-driven. Thus, the agnostic bounds serve as the most robust baseline.

\section{Comparison with previous analyses}

\begin{figure}
    \centering
    \vspace{-10pt}
    \includegraphics[width=0.49\columnwidth]{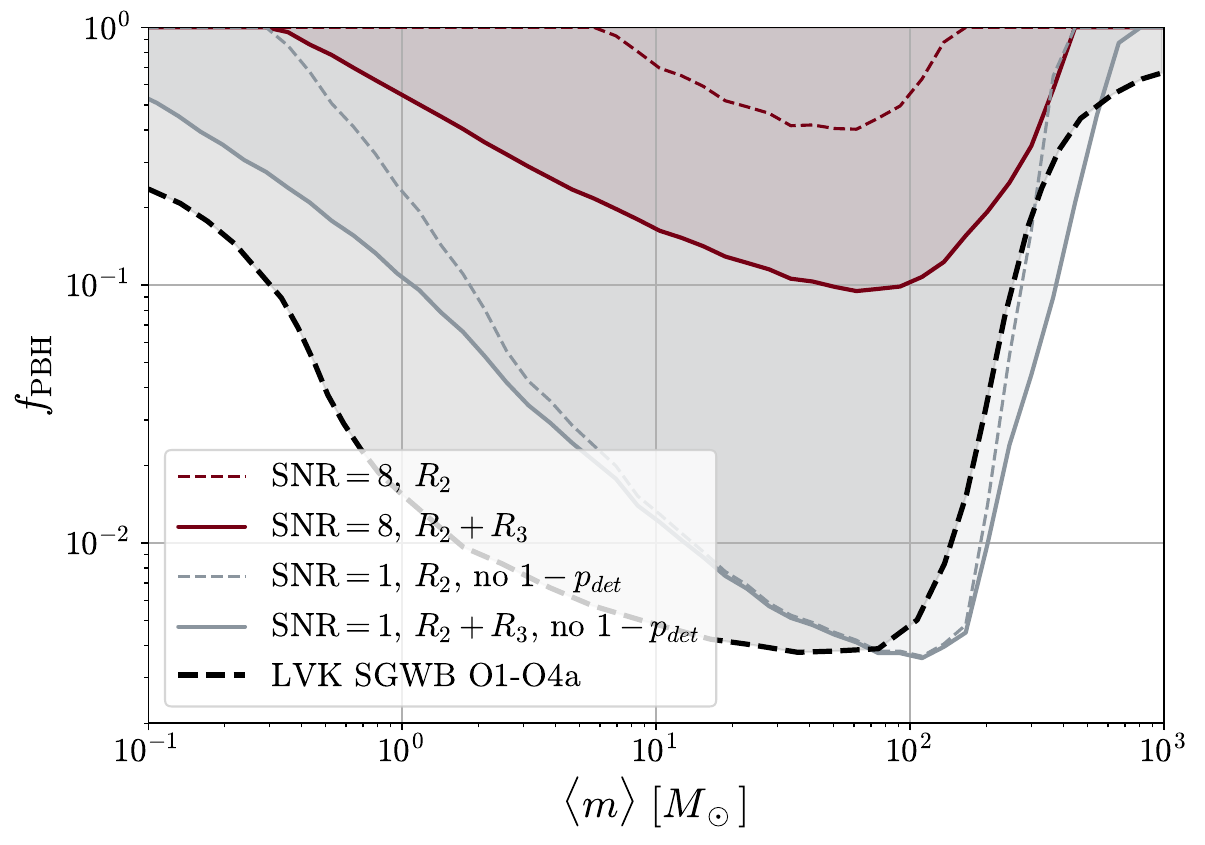}
    \includegraphics[width=0.49\columnwidth]{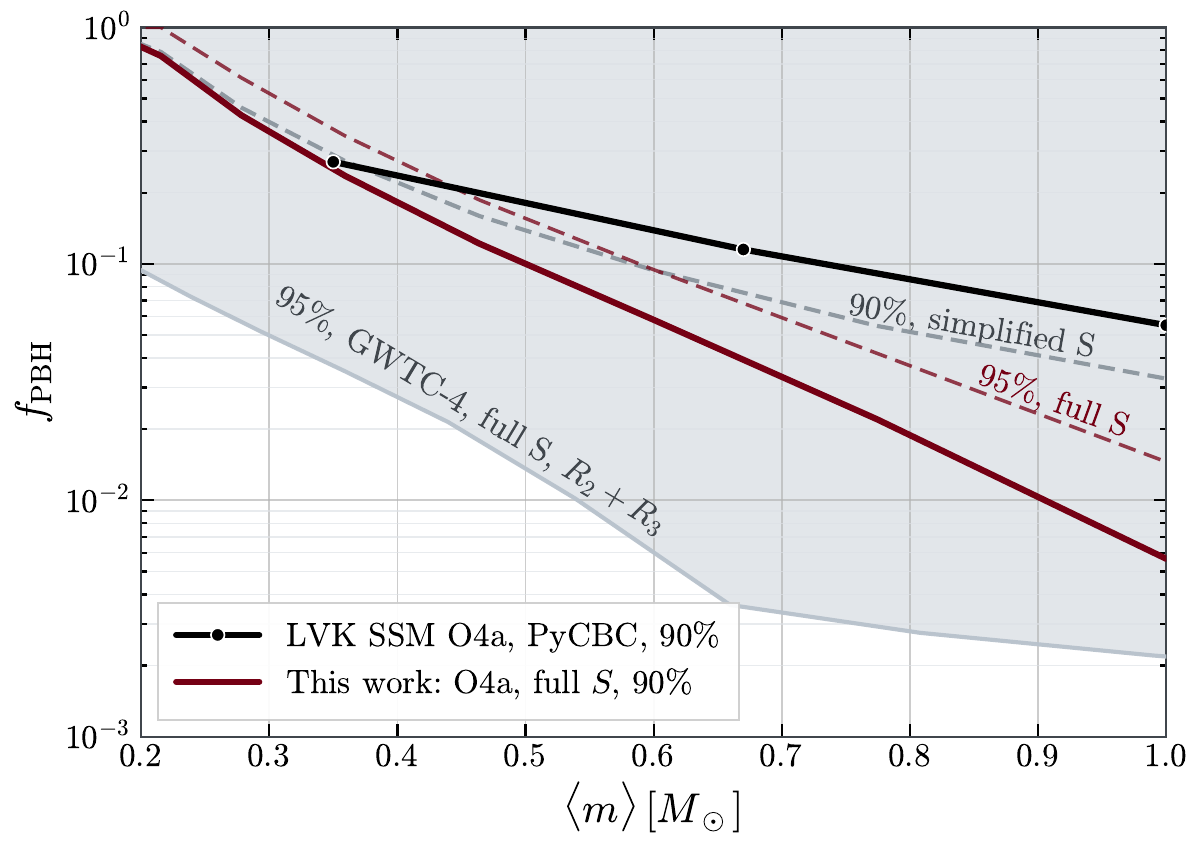}
    \caption{ \emph{Left panel:} Comparison of our SGWB constraint (gray solid) with the constraint derived in Ref.~\cite{LIGOScientific:2025kry} (black dashed). The solid (and dotted) curves show our constraints with (and without) the three-body channel. The gray curves show our most optimistic constraint with SNR=1 and neglecting the $1-p_{\rm det}$ factor. \emph{Right panel:} Comparison of our constraint (gray solid) with the 90\% CL constraint derived in Ref.~\cite{LIGOScientific:2026wxz} (black solid). The red solid and dashed curves show our 90\% and 95\% CL constraints with the three-body channel neglected but the full two-body suppression factor is retained. The gray dashed curve shows the 90\% CL constraint obtained using neither the three-body channel nor the full two-body suppression factor.}
    \label{fig:Comparison}
\end{figure}

Recently, Ref.~\cite{LIGOScientific:2026wxz} derived an upper limit of $\fpbh \lesssim \mathcal{O}(0.1)$ in the $0.4-1\,\msun$ mass interval for monochromatic PBH mass functions. These bounds are weaker by approximately an order of magnitude than those presented in this work. A direct comparison is shown in the right panel of Fig.~\ref{fig:Comparison}. The discrepancy can be mostly attributed to differences adopted in the underlying PBH binary formation model: the analysis of Ref.~\cite{LIGOScientific:2026wxz} employed a simplified suppression factor for the two-body formation channel and did not include the three-body channel, which is expected to dominate for abundances $\fpbh \gtrsim 0.04$, where the two-body channel becomes strongly suppressed. In Fig.~\ref{fig:Comparison}, these effects account for most of the difference between the results. The remaining discrepancy is relatively minor and likely arises from differences in the analysis methodology. The analysis in Ref.~\cite{LIGOScientific:2026wxz} estimated the sensitive hypervolume using injection campaigns and false-alarm-rate-based detection criteria, whereas our analysis relies on representative detector noise curves and a fixed SNR threshold. However, Fig.~\ref{fig:Comparison} shows that the grey dashed and black solid curves differ only by $\mathcal{O}(10\%)$ in the resulting constraint on $f_{\rm PBH}$.

The bounds from SGWBs produced by PBH binaries in Ref.~\cite{LIGOScientific:2025kry} are significantly stronger than what we found in Fig.~\eqref{fig:LimitsO4a_Combined}. This discrepancy cannot be explained by differences in the PBH binary population modelling since Ref.~\cite{LIGOScientific:2025kry} considers only the two-body formation channel with the same suppression factors used here. Therefore, weaker constraints are expected since the three-body channel is not included. As illustrated in the left panel of Fig.~\ref{fig:Comparison}, this mismatch persists even under the most optimistic assumptions in our framework, where we set ${\rm SNR}_{\rm thr} = 1$ and omit the $1-p_{\rm det}$ factor that excludes the contribution of individually resolvable sources from the background.

Finally, searches for planetary-mass binaries of Ref.~\cite{LIGOScientific:2025vwc} reported a constraint on the PBH abundance from asymmetric mass-ratio binaries. However, because their analysis assumes a broad mass function encompassing both planetary- and solar-mass black hole binaries, a direct comparison with our results is not meaningful. In agreement with us, Ref.~\cite{LIGOScientific:2025vwc} does not constrain planetary-mass PBHs in the equal-mass case.

\section{Conclusions}

Using LVK data up to the O4a run, we have derived updated constraints on the PBH abundance from both the observed population of binary mergers and the non-observation of a SGWB. The limits from individually resolvable binaries provide the most stringent constraints to date for monochromatic mass functions in the range $0.6~\msun \lesssim \langle m \rangle \lesssim 100~\msun$. For broader mass functions, the constraints weaken slightly near $\langle m \rangle \lesssim 1~\msun$ and $\sim 100~\msun$, but extend to cover masses up to $\langle m \rangle \sim 10^3~\msun$. The SGWB non-detection yields roughly three orders of magnitude weaker bounds that are more sensitive to the mass function width than the resolvable binary limits.

We have explored a possible population of galactic subsolar binaries, finding that the current LVK sensitivity does not yet constrain such systems. Nevertheless, we have identified a sensitivity floor in the mass range $10^{-4}~\msun \lesssim \langle m \rangle \lesssim 10^{-3}~\msun$, where LVK could probe PBH fractions down to $f_{\rm PBH} \approx 0.2$. Overall, we found that LVK can probe PBH scenarios in the mass range $10^{-4}-10^4 \msun$.

To derive PBH limits that are agnostic about the ABH population, we have introduced two independent, data-driven methods based on the observed binary sample. These methods yield consistent results, modestly relaxing the constraints in the mass range $2~\msun < \langle m \rangle < 20~\msun$ while providing robust, ABH model independent bounds. As a third alternative, we have performed joint fits to the ABH and PBH populations, where the ABH population is described by a generic parametrisation capable of covering a large range of ABH models. Although generic, this approach allows us to constrain ABHs models by imposing astrophysically motivated limits on the parameter space. We found that allowing the ABH low mass cutoff to vary freely and retaining all events can spuriously favour a PBH component ($\ln B \simeq 30$), while fixing $m_{\min}=5~\msun$ and excluding the low mass gap outlier removes any significant preference ($\ln B \simeq 1.7$) for PBHs. Overall, our bounds are independent of assumptions about the astrophysical BH population and represent the most stringent constraints to date.

\section*{Acknowledgments}
This document has received an internal LIGO DCC number of P2500675. This work was supported by Estonian Research Council grants PSG869, TARISTU24-TK3 and TARISTU24-TK10 and the Center of Excellence program TK202. The work of V.V. was supported by the European Union's Horizon Europe research and innovation program under the Marie Sk\l{}odowska-Curie grant agreement No. 101065736.This work is partially supported by the Spanish MCIN/AEI/10.13039/501100011033 under the Grants No. PID2020-113701GB-I00 and PID2023-146517NB-I00, some of which include ERDF funds from the European Union, and by the MICINN with funding from the European Union NextGenerationEU (PRTR-C17.I1) and by the Generalitat de Catalunya. IFAE is partially funded by the CERCA program of the Generalitat de Catalunya. This material is based upon work supported by NSF’s LIGO Laboratory, which is a major facility fully funded by the National Science Foundation and operates under Cooperative Agreement No. PHY-1764464.
This research has made use of data, software, and/or web tools obtained from the Gravitational Wave Open Science Center (\url{https://www.gw-openscience.org/}), a service of the LIGO Laboratory, the LIGO Scientific Collaboration, and the Virgo Collaboration.

\appendix

\section{PBH binary population models}
\label{app:PBHBs}

The early two-body PBH binary formation channel results in the merger rate~\cite{Ali-Haimoud:2017rtz,Raidal:2018bbj,Vaskonen:2019jpv,Raidal:2024bmm}
\be\label{eq:R2}
    \frac{\td \Rate_{\mathrm{\rm PBH},2}}{\td m_1 \td m_2}
    = \frac{1.6\times 10^6}{\text{Gpc}^3\text{yr}^1}
    \fpbh^{\frac{53}{37}}\left[ \frac{t}{t_0}\right]^{-\frac{34}{37}}\left[ \frac{M}{\msun}\right]^{-\frac{32}{37}}  
    \eta^{-\frac{34}{37}} S  \left[ \psi,\fpbh,M \right]\frac{\psi(m_1)\psi(m_2)}{m_1 m_2}\, ,
\ee
where $\fpbh\equiv \rho_{\mathrm{\rm PBH}}/\rho_{\mathrm{DM}}$ is the total fraction of DM in the form of PBHs, $t_0$ the current age of the Universe, $M=m_1+m_2$ the total mass of the binary system, $\eta=m_1m_2/M^2$ its symmetric mass ratio, $S$ the suppression factor, $\psi(m) \equiv \rho_{\rm PBH}^{-1}\td \rho_{\rm PBH}/\td \ln m$ the PBH mass function. To evaluate the suppression factor, we use the analytical prescriptions of~\cite{Hutsi:2020sol}. 

The PBH merger rate from the early three-body channel results can be estimated as~\cite{Vaskonen:2019jpv, Raidal:2024bmm}
\bea\label{eq:R3}
    \frac{\td R_{\rm PBH,3}}{\td m_1 \td m_2}
    &\approx \frac{7.9 \times 10^4}{\rm Gpc^{3}\,yr } 
    \left[ \frac{t}{t_0}\right]^{\frac{\gamma}{7} - 1}
    f_{\rm PBH}^{\frac{144 \gamma}{259}+\frac{47}{37}}
    \left[ \frac{\langle m \rangle}{\msun}\right]^{\frac{5 \gamma -32}{37}}\!
    \\ 
    &\times
    \left(\frac{M}{2\langle m \rangle}\right)^{\frac{179 \gamma }{259}-\frac{2122}{333}}\!\!\!
    (4\eta)^{-\frac{3 \gamma }{7}-1}  
    \mathcal{K} \,
    \frac{e^{-3.2 (\gamma - 1)}\gamma}{28/9-\gamma}
    \bar{\mathcal{F}}(m_1,m_2)
    \frac{\psi(m_1)\psi(m_2)}{m_1 m_2}\,,
\eea
where $\langle m \rangle$ is the mean mass of the distribution, $\gamma \in [1,2]$ is a parameter that describes the dimensionless distribution of angular momenta $j$ (when assumed to take the form $P(j) = \gamma j^{\gamma-1}$), the factor $\bar{\mathcal{F}}(m_1,m_2)$ is a correction that accounts for the composition of masses in the initial 3-body system, assuming that the lightest PBHs is ejected, and $\mathcal{K}$ is a term that accounts for the hardening of the early binary in encounters with other PBHs. We use the expression for $\bar{\mathcal{F}}(m_1,m_2)$ from Ref.~\cite{Raidal:2024bmm}. The value for the parameter $\gamma$ controls the angular momentum distributions, with $\gamma=2$ corresponding to the equilibrium distribution, while a super-thermal distribution would lead to $\gamma = 1$~\cite{Raidal:2018bbj,2019Natur.576..406S}. We take $\gamma=1$ in this work. Similarly, we choose $\mathcal{K} = 4$ as suggested by numerical simulations~\cite{Raidal:2018bbj}.

We remark that the two-body channel merger rate with initial clustering was recently derived in~\cite{Kasai:2026kpv} for monochromatic mass functions. However, since the strength of clustering is limited by other observations~\cite{DeLuca:2022uvz} and because more work is needed on extended mass functions and the three-body channel to fully model PBH binary populations with initial clustering, we will not consider these effects in this work. Ref~\cite{Kasai:2026kpv} also imposed cuts on the initial distance of the PBH pair, which are not included in the merger rate~\eqref{eq:R2}. We checked that such cuts do not significantly alter our results.

\section{Hierarchical Bayesian analysis}
\label{app:methods}

To evaluate the likelihood for resolvable binaries, Eq.~\eqref{eq:hierarchicalBayesian}, we implement our models on top of the \texttt{ICAROGW 2.0} code, as it already contains optimised functions to evaluate this expression on GPUs~\cite{Mastrogiovanni:2023zbw}. We rely on the posterior samples provided by LVK in the different data releases~\cite{LIGOScientific:2019lzm,KAGRA:2023pio,LIGOScientific:2025snk}. We link each of the $N_p$ posterior samples, using Bayes' theorem, to the individual likelihood as $p(\theta|d_i,\Lambda)\propto\mathcal{L}(d_i|\theta)\pi_{\varnothing}(\theta|\Lambda)$, where $\pi_{\varnothing}(\theta|\Lambda)$ represents the prior used for the initial parameter estimation of such an event. Then, the integral term of each event can be approximated using Monte-Carlo integration as~\cite{Mastrogiovanni:2023zbw,LIGOScientific:2020kqk}
\be \label{eq:Lint}
    \int\!\mathcal{L}(d_i|\theta)\pi(\theta|\Lambda)\td \theta 
    \!\approx\!\frac{1}{N_p}\!\sum_{j=1}^{N_p}\!\frac{\pi(\theta_{ij}|\Lambda)}{\pi_{\varnothing}(\theta_{ij}|\Lambda)}
    \!\equiv\!\frac{1}{N_p}\!\sum_{j=1}^{N_p}w_{ij} \,.
\ee
In order to ensure numerical stability during this process, the effective number of posterior samples per event, defined as~\cite{Mastrogiovanni:2023zbw,Talbot:2023pex,LIGOScientific:2021aug}
\be
    N_{\mathrm{eff},i} = \bigg[  \displaystyle\sum_j^{N_p} w_{ij} \bigg]^2 \!\bigg/\, \displaystyle\sum_j^{N_p} w_{ij}^2 \,,
\ee
is set to $20$. For each event, we select a random subsample of $N_p = 2048$ posterior samples to evaluate Eq.~\eqref{eq:Lint}. 

The other relevant quantity to be computed is the number of expected events. We follow the same prescription of using Monte Carlo integration, and we evaluate Eq.~\eqref{eq:Nexpected} using $N_g$ generated samples from an injection set as~\cite{Mastrogiovanni:2023zbw}
\be
    N(\Lambda)\approx \frac{T}{N_g}\sum_{j= 1}^{N_{\mathrm{det}}}\frac{\pi(\theta_j|\Lambda)}{\pi_{\mathrm{inj}}(\theta_j)}\equiv \frac{T}{N_g}\sum_{j=1}^{N_{\mathrm{det}}}s_j\, ,
\ee
where $\pi_{\mathrm{inj}}(\theta_j)$ is the prior probability of the $j$th injection. As for the posterior, a numerical stability estimator can be defined for injections as \cite{Farr:2019rap,Mastrogiovanni:2023zbw}
\be
    N_{\mathrm{eff,inj}} = \frac{\bigg[\displaystyle \sum_j^{N_{\mathrm{det}}} s_j \bigg]^2}{\displaystyle\sum_j^{N_{\mathrm{det}}} s_j^2-\displaystyle\frac{1}{N_{g}}\bigg[\displaystyle\sum_j^{N_{\mathrm{det}}} s_j\bigg]^2}\, ,
\ee
which we set to $N_{\mathrm{eff,inj}}>4\Nobs$~\cite{Farr:2019rap}. Finally, the log-likelihood is evaluated as~\cite{Mastrogiovanni:2023zbw}
\be
    \ln\!\mathcal{L}(\{d\}|\Lambda) 
    \!\approx\!-\frac{T_\mathrm{obs}}{N_g}\sum_{j}^{\Ndet}s_j +\!\sum_{i}^{\Nobs}\ln\!\left[ \frac{T_\mathrm{obs}}{N_{p}}\sum_j^{N_p}w_{ij} \right]\!.
\ee

To generate the injection set, we follow the semi-analytical procedure explained in Refs.~\cite{Vaskonen:2019jpv,Andres-Carcasona:2024wqk}, taking into account the various duty cycles of the different observing runs.

\section{Agnostic treatment of ABHs}
\label{app:agnosticABHs}

The most conservative constraints on the PBH abundance are obtained without restricting which subset of the observed events is of primordial origin. Such constraints can be derived by systematically selecting different subsets of the observed events and fitting each subset with a PBH population model. The resulting upper bound on $f_{\rm PBH}$ is then taken from the subset that yields the weakest (i.e., least stringent) constraint within the class of models considered. In this way, the inferred constraint is completely independent of any modelling of the remaining, non-primordial events, which are typically described using a phenomenological, astrophysically motivated parametrisation.

However, since the number of possible subsets that can be constructed from a given set of events increases exponentially with the total number of events, performing dedicated PBH fits for all such subsets is computationally intractable. Consequently, it is necessary to employ an optimised or approximate strategy to explore the space of subsets. A specific approach to this optimisation will be presented below.

As a starting point, we will first consider a decomposition of the likelihood \eqref{eq:hierarchicalBayesian} in case there are two distinct channels of BH binary formation. We will denote these channels A and B, respectively (e.g., A=PBH and B=ABH). A decomposition of the merger rate $\td \Rate = \td \Rate_A + \td \Rate_B$ implies that the hyperprior can be split as
\be\label{eq:pi_AandB}
    \pi(\theta|\Lambda) = \pi_A(\theta|\Lambda_A) + \pi_B(\theta|\Lambda_B)\,,
\ee
where $\Lambda = \Lambda_A \cup \Lambda_B$. The decomposition of everything else then follows from Eq.~\eqref{eq:pi_AandB}: we also have $N(\Lambda) = N_A(\Lambda_A) + N_B(\Lambda_B)$ and the likelihood \eqref{eq:hierarchicalBayesian} can be recast as
\bea\label{eq:hierarchicalBayesianAB}
    \mathcal{L}(\{d\}|\Lambda) 
    &\propto e^{-N_A(\Lambda) - N_B(\Lambda)}\prod_{i=1}^{\Nobs} (\ell_{i,A} + \ell_{i,B})
    \\
    &= \sum_{S_A \subset S} 
    \mathcal{L}(\{d\}_{S_A}|\Lambda_A)
    \mathcal{L}(\{d\}_{S\setminus S_A}|\Lambda_B)\,,
\eea
where the sum runs over all subsets $S_A$ of the set of events $S$ and we define the per-event marginal likelihood in the model $M \in \{A,B\}$
\be
    \ell_{i,M} = \int \mathcal{L}(d_i|\theta)\pi_{M}(\theta|\Lambda)\td \theta \, .
    \label{eq:individual_likelihood}
\ee
and we define the channel $M$ likelihood for a subset of $S_M \subset S$ events $\{d\}_{S_M} \equiv \{d_i: i \in S_M\}$ as
\be
    \mathcal{L}(\{d\}_{S_M}|\Lambda_M) \equiv e^{-N_M(\Lambda)}\prod_{i \in S_M} l_{i,M}\,.
\ee
The reformulation in Eq.~\eqref{eq:hierarchicalBayesianAB} makes it explicit how the likelihood distributes all events between models $A$ and $B$: it considers all possible ways of dividing the events between different channels. Since testing all of them is computationally unfeasible, we tackle this problem with two approaches.

\begin{figure}
    \centering
    \includegraphics[width=0.6\columnwidth]{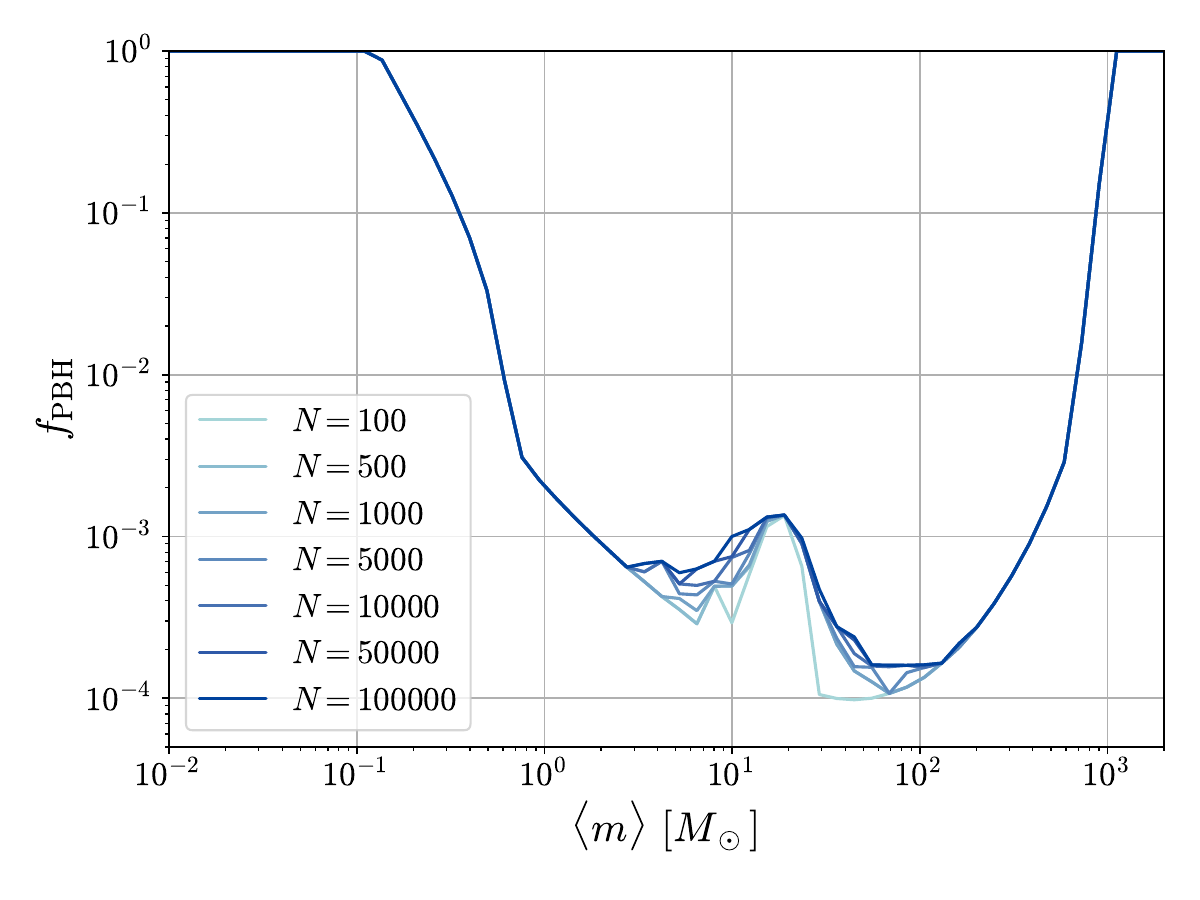}
    \caption{Agnostic limits obtained via the random subset method for a log-normal mass function with $\sigma=0.6$, for an increasing number of combinations.}
    \label{fig:Convergence_agnostic}
\end{figure}

The first is to test only some of the subsets chosen purely at random. This method has already been used in Refs.~\cite{Hutsi:2020sol,Andres-Carcasona:2024wqk} to quote agnostic limits to the PBH population. The more combinations that are tested, the more likely it is to add one that is less constraining than the rest at a given mass-region. For a sufficiently large number of subsamples tested, the result reaches a plateau, and we can consider that to be a reasonable approximation of the actual agnostic limits. Fig.~\ref{fig:Convergence_agnostic} shows the constraint obtained for different values of the subsamples, $N$, added. Since this procedure is cumulative, the results for a larger subsample can only be equal to or less restrictive than those from the previous ones. This implies that this method produces conservative limits.

The second method is to slightly modify the likelihood so that some events are automatically discarded. The modified log-likelihood reads as 
\be
    \ln \mathcal{L}_\alpha = -N(\Lambda) + \sum_{i=1}^{\Nobs} \ln \left(\alpha + \ell_i(\Lambda)\right ) \, .
    \label{eq:lnL_alpha}
\ee

This modified version of the likelihood presented in Eq.~\eqref{eq:hierarchicalBayesian} allows us to set a threshold that accounts for only some events, governed by the value of $\alpha$. In the limiting case where $\alpha\to0$, the original likelihood is recovered, which corresponds to the case where all detected events are considered primordial. On the other hand, if $\alpha \to \infty$, then this term becomes larger than the likelihood of the event, making the sum over the events simply a constant. Therefore, $\ln \mathcal{L}_\alpha \propto -N(\Lambda)$ becomes the likelihood in which none of the events is primordial. In the range $\alpha \in [0, \infty)$, we can obtain a set of constraints between these two limiting cases. The role of $\alpha$ is to introduce a soft floor to the contribution of each event.  To see this, note that the score with respect to any hyperparameter component $\Lambda_j$ is
\begin{align}
\frac{\partial \ln \mathcal{L}_{\alpha}}{\partial \Lambda_j}
 &=
-\frac{\partial N}{\partial \Lambda_j}
+\sum_{i=1}^{N_{\rm obs}}
\frac{1}{\alpha+\ell_i(\Lambda)}\frac{\partial \ell_i(\Lambda)}{\partial \Lambda_j}
\nonumber \\&=
-\frac{\partial N}{\partial \Lambda_j}
+\sum_{i=1}^{N_{\rm obs}}
w_i(\alpha,\Lambda)\frac{\partial \ln \ell_i(\Lambda)}{\partial \Lambda_j}\, ,
\label{eq:dLdLambda}
\end{align}
where the per–event weight is 
\be
    w_i(\alpha,\Lambda) \equiv \frac{\ell_i(\Lambda)}{\alpha+\ell_i(\Lambda)}\in[0,1]\, .
\ee

Thus, events for which the PBH population predicts large $\ell_i$ (relative to $\alpha$) are effectively kept with a weight $\approx 1$, whereas events with small $\ell_i\ll\alpha$ are smoothly down–weighted ($w_i\approx 0$).  Equivalently, this modified likelihood implements a continuous relaxation of a hard selection of the events, with the two limiting cases already discussed. It is useful to define the effective number of PBH–like events
\be
    N_{\rm eff}(\alpha,\Lambda)\equiv\sum_{i=1}^{N_{\rm obs}} w_i(\alpha,\Lambda)\in[0,N_{\rm obs}]\, ,
\ee
which provides a direct intuition for how much of the catalogue the model is using.  As $\alpha$ increases, $N_{\rm eff}$ decreases monotonically from $N_{\rm obs}$ to $0$. The construction of this likelihood can be viewed as a robust, ABH–agnostic mixture where each event arises either from the PBH population with marginal $\ell_i(\Lambda)$ or from any other channel with an unknown per–event marginal likelihood that we approximate by an event–independent scale, $\alpha$.

Although the random subsampling method can be interpreted as a hard gate where an event contributes $\ln \ell_i$ or nothing, this other likelihood replaces that step function with smooth weights, yielding a differentiable function with stable optimisation and sampling, while preserving the same limiting behaviours. In practice, we observe a similar behaviour of constraints because events that are inconsistent with the PBH population are automatically down–weighted, and those well–described by $\Lambda$ contribute almost as in the standard hierarchical likelihood.

The connection to the random–subset method is exact under a simple agnostic mixture. Let binary selectors $s_i\in\{0,1\}$ indicate whether the event $i$ is treated as primordial, and suppose that those with a non–PBH origin contribute some constant $C>0$ at the product level. In this case, the likelihood becomes 
\be
    \mathcal{L}(\Lambda|s) =e^{-N(\Lambda)}\prod_{i=1}^{N_{\rm obs}}\left[ \ell_i(\Lambda) \right]^{s_i} C^{1-s_i}\, .
\ee

If $s_i\sim{\rm Bernoulli}(\pi)$ i.i.d., corresponding to $p(s)=\prod_i\pi^{s_i}(1-\pi)^{1-s_i}$, we can marginalise over all subsets to obtain the mixed likelihood as
\bea
    \mathcal{L}_{\rm mix}(\Lambda\mid \pi)
    &= \sum_s p(s)\mathcal{L}(\Lambda|s,C)
    \\  
    &=e^{-N(\Lambda)}\prod_{i=1}^{N_{\rm obs}}\big(\pi\,\ell_i(\Lambda)+(1-\pi)\,C\big)\, ,
\eea
with the sum in the first row running over all possible assignments of $\{0,1\}$ for $s_i$.
Up to an irrelevant constant, this is identical to $\mathcal{L}_{\alpha}$ after reparametrisation
\be
\alpha \;=\; \frac{1-\pi}{\pi}\,C,
\qquad
w_i(\alpha,\Lambda)
=\frac{\pi\,\ell_i}{\pi\,\ell_i+(1-\pi)C}.
\ee
Hence, the likelihood in Eq.~\ref{eq:lnL_alpha} is the closed form of random subset marginalisation for this mixture. We therefore refer to this likelihood as the subset marginalised likelihood (SML) used in the main text. 

Under this mixture, the posterior probability that an event is primordial is simply 
\be
    \mathbb{P}(s_i= 1|\Lambda,\pi) = \frac{\pi\,\ell_i}{\pi\,\ell_i+(1-\pi)C} = w_i (\alpha,\Lambda)\,,
\ee
which can be linked to the effective number of PBH events as
\be
    N_{\rm eff}(\alpha,\Lambda)=\sum_{i=1}^{N_{\rm obs}} w_i(\alpha,\Lambda) = \sum_{i=1}^{N_{\rm obs}}\mathbb{P}(s_i= 1|\Lambda,\pi) \, ,
\ee
allowing us to interpret $N_{\rm eff}$ as the posterior of the expected number of PBHs.

A practical difference arises when one averages log–likelihoods over random subsets. By Jensen’s inequality,
\be
\mathbb{E}\!\left[\ln\mathcal{L}(\Lambda\mid\mathbf{s})\right]
\;\le\; \ln \mathbb{E}\!\left[\mathcal{L}(\Lambda\mid\mathbf{s})\right]
=\ln \mathcal{L}_{\rm mix}(\Lambda\mid \pi,C),
\ee
so log–averaging subsets produces a downward bias in $\ln\mathcal{L}$ (i.e., more conservative posteriors) relative to the closed form. Regarding the effect on the bounds, consider a PBH parameter $\Lambda_j$, from Eq.~\eqref{eq:dLdLambda} one sees that introducing $\alpha>0$ uniformly attenuates the informative event terms by the $w_i\in(0,1)$ while leaving the Poisson penalty $-\partial N/\partial\Lambda_j$ unchanged. Therefore, the posterior mass shifts toward regions favoured by the count term, typically yielding higher (more conservative) upper limits on $\Lambda_j$ unless many events have $\ell_i\gg\alpha$.

Bias enters if the mixture is misspecified. If the agnostic contribution is heterogeneous, $\alpha=\alpha_i$, but we fit a single $\alpha$, the fitted log–likelihood differs by
\be
    \Delta \ell(\Lambda)
    =\sum_i \Big[\ln(\alpha+\ell_i)-\ln(\alpha_i+\ell_i)\Big].
\ee
For small $\delta\alpha_i=\alpha-\alpha_i$,
\be
    \Delta \ell(\Lambda)\approx
    \sum_i \frac{\delta\alpha_i}{\alpha_i+\ell_i}
    -\sum_i \frac{\delta\alpha_i}{\alpha_i+\ell_i}\,
    \frac{\partial \ln \ell_i}{\partial \Lambda}^{\!\top}\delta\Lambda,
\ee
and a standard $M$–estimation expansion gives the leading hyperparameter bias
\bea
    \delta\Lambda 
    &\approx
    \mathcal{I}(\Lambda)^{-1}
    \sum_i \frac{\delta\alpha_i}{(\alpha_i+\ell_i)^2}\,\frac{\partial \ell_i}{\partial \Lambda}\,,
    \qquad\qquad
    \mathcal{I}(\Lambda)
    \equiv
    -\,\frac{\partial^2 \ln\mathcal{L}_{\alpha}}{\partial \Lambda\,\partial \Lambda^{\!\top}}\,.
\eea
Overestimating $\alpha$ for events in regions where $\partial \ell_i/\partial\Lambda$ points toward larger PBH abundance pushes the estimate toward smaller abundances (weaker $\ell_i$), i.e., towards stronger down–weighting of PBH–like regions and consequently yielding weaker limits.

We scan $\alpha$ on a logarithmic grid over $\alpha\geq 0$, including the limiting case $\alpha=0$, and at each characteristic mass report the largest upper limit on $f_{\rm PBH}$ obtained across the scan. The upper end of the range is increased until this conservative envelope has converged to its large-$\alpha$ limit.

\section{Parametric treatment of ABHs and the ABH+PBH fits}
\label{app:parametrizedABHs}

We perform the combined fit of ABH and PBH populations using the hierarchical Bayesian likelihood~\eqref{eq:hierarchicalBayesian} with
\be
    \frac{\td \Rate}{\td m_1 \td m_2}(\theta|\Lambda) \!=\! \frac{\td \Rate_ {\rm ABH}}{\td m_1 \td m_2}(\theta|\Lambda_{\rm ABH}) + \frac{\td \Rate_ {\rm PBH}}{\td m_1 \td m_2}(\theta|\Lambda_{\mathrm{\rm PBH}})\,,
\ee
where $\Lambda = \Lambda_{\rm ABH}\cup \Lambda_{\rm PBH}$. We employ a power-law + 2 PEAK model (PL+2P, also named MULTI PEAK in some references) for parametrising the astrophysical population\footnote{We employ this model instead of the power-law + peak model used in previous works (see e.g.~\cite{Andres-Carcasona:2024wqk}), because the amount of statistics acquired through the O4a observing run of LVK is enough to start seeing a secondary peak in the mass spectrum with enough statistical significance. For example, Ref.~\cite{LIGOScientific:2025pvj} already pointed out a decisive Bayes factor in favour of the two peaks model instead of a single one.}~\cite{Mastrogiovanni:2023zbw}:
\be
    \frac{\td \Rate_ {\rm ABH}}{\td m_1 \td m_2}(\theta|\Lambda_{\rm ABH}) = R_0\pi^{m_1}(m_1)\pi^{m_2}(m_2|m_1)\pi^{z}(z)\, ,
\ee
where for the redshift evolution, we consider a power-law of the form 
\be
    \pi^z(z) = (1+z)^\kappa\, .
\ee 
The probability distribution of the primary mass is
\begin{align}
    \pi^{m_1}(m_1)=&[(1-\lambda)\mathcal{P}(m_1|m_{\min},m_{\max},-\alpha)\nonumber \\
    &+\lambda\lambda_{\rm low}\mathcal{G}(m_1|\mu_{\rm low},\sigma_{\rm low}) \nonumber \\
    &+\lambda(1-\lambda_{\rm low})\mathcal{G}(m_1|\mu_{\rm high},\sigma_{\rm high})]\nonumber \\ &\times \mathcal{S}(m_1|m_{\min},\delta_m)\, ,
\end{align}
where $\mathcal{P}$ and $\mathcal{G}$ are normalised truncated power law and Gaussian distributions, respectively, and $\mathcal{S}(m_1|\delta_m,\mmin)$ is a smoothing function:
\bea 
    &\mathcal{S}(m_1|\mmin,\delta_m) = \\ &\begin{cases} 
      0\,, & m< \mmin \\
      [f(m-\mmin,\delta_m)+1]^{-1}, & \mmin\leq m< \mmin+\delta_m \\
      1\,, & m\geq \mmin+\delta_m
    \end{cases}
\eea
with
\be
    f(m',\delta_m) = \exp\left( \frac{\delta_m}{m'}+\frac{\delta_m}{m'-\delta_m}\right) \,.
\ee
The probability distribution of the secondary mass is described by a smoothed truncated power law conditioned on the primary mass:
\be
    \pi^{m_2}(m_2|m_{\min},m_1,\beta) = \mathcal{P}(m_2|\mmin,m_1,\beta) \mathcal{S}(m_2|\delta_m,\mmin) \,.
\ee
The ABH parametrisation is described by 13 parameters: 
\be
        \Lambda_{\rm ABH} = \{ R_0,\alpha,\beta, \mmin, \mmax, \delta_m, \lambda, \mu_{\rm high}, \sigma_{\rm high}, \lambda_{\rm low}, \mu_{\rm low}, \sigma_{\rm low}, \kappa \}\, .
\ee

\begin{figure*}
    \centering
    \includegraphics[width=\textwidth]{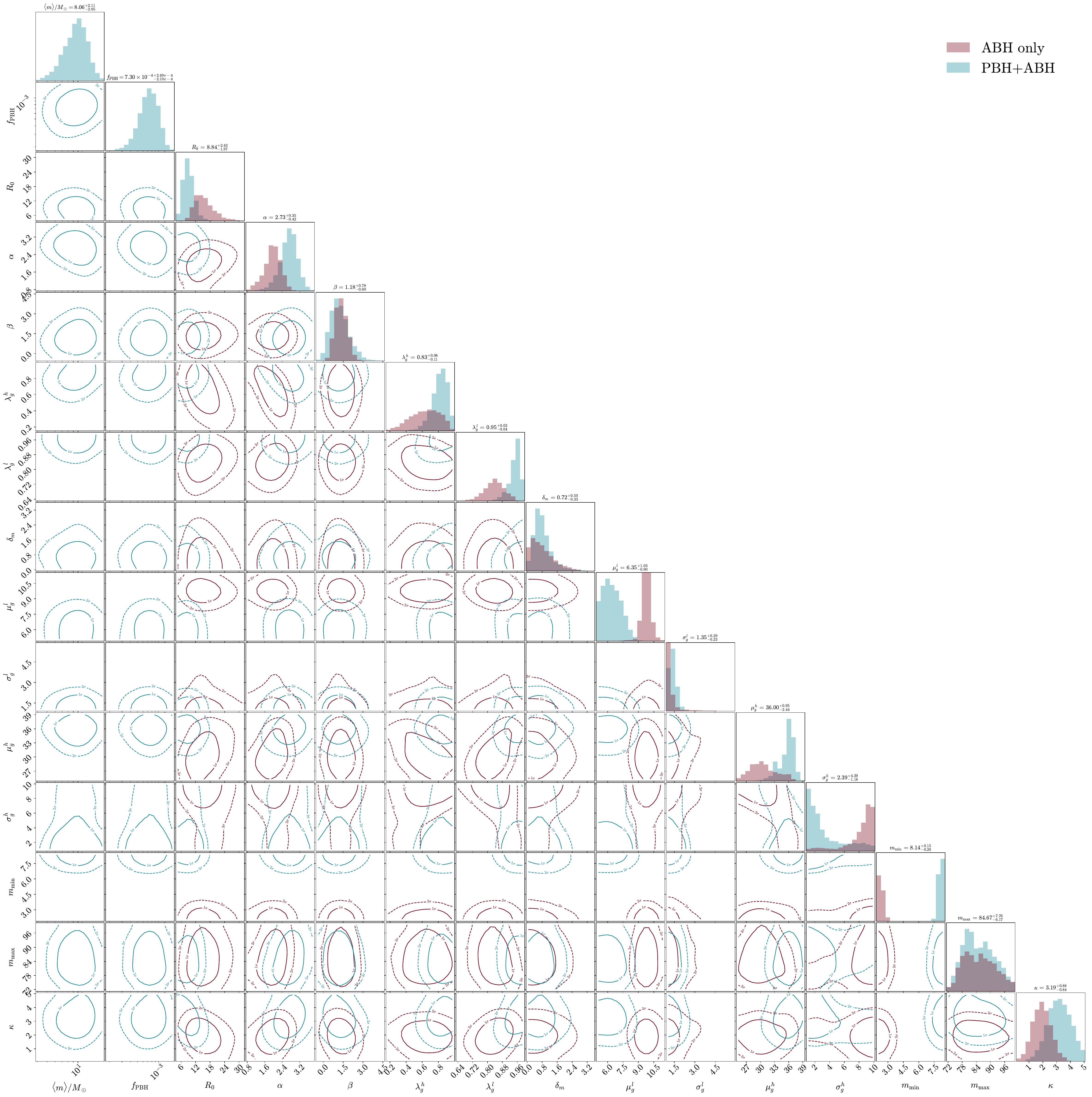}
    \caption{Posteriors for the BH binary merger rate model combining PBH and ABH binaries (blue) and for the model containing only ABH binaries (red). The PBH binaries are assumed to have a log-normal mass function with width $\sigma = 0.6$ and the ABH binaries are described by a phenomenological PL+2P model. $R_0$ in Gpc$^{-3}$yr$^{-1}$.}
    \label{fig:CornerPlot}
\end{figure*}

\begin{table}
	\centering
	\begin{tabular}{ccc}
		\hline \hline
		\textbf{Parameter} & \textbf{Prior} & \textbf{Units}\\
		\hline \hline \\[-9pt]
            $R_0$ & LogU$(10^{-2},10^3)$ & Gpc$^{-3}$yr$^{-1}$\\ 
		$\alpha$ & U$(-4,12)$ & - \\ 
            $\beta$ & U$(-4,12)$ & - \\ 
            $m_{\min}~^*$ & U$(2,10)$ & $\msun$ \\ 
            $m_{\max}$ & U$(40,100)$ & $\msun$ \\ 
            $\delta_m$ & U$(0,10)$ & $\msun$ \\ 
            $\mu_{\rm high}$ & U$(25,60)$ & $\msun$ \\ 
            $\sigma_{\rm high}$ & U$(1,10)$ & $\msun$ \\ 
            $\lambda$ & U$(0,1)$ & - \\
            $\mu_{\rm low}$ & U$(5,20)$ & $\msun$ \\ 
            $\sigma_{\rm low}$ & U$(1,10)$ & $\msun$ \\ 
            $\lambda_{\rm low}$ & U$(0,1)$ & - \\
            $\kappa$ & U$(0,5)$ & - \\
            $f_{\rm PBH}$ & LogU$(10^{-5},1)$ & - \\
            $m_c$ & LogU$(10^{-2},2\times10^3)$ & $\msun$ \\
            \hline\hline
	\end{tabular}
      \caption{Priors used for the PBH+ABH fit. $^*$ This variable might be fixed in some of the fits, we specify it in the text or caption of such cases. }
      \label{tab:priors}
\end{table}

\begin{table}
	\centering
	\begin{tabular}{@{\extracolsep{0.5cm}}ccc}
		\hline \hline
		\textbf{Scenario} & $\pmb{\ln\mathcal{Z}}$ & $\pmb{\ln\mathcal{B}}$\\
		\hline \hline
            \multicolumn{3}{c}{All events and free $\mmin$} \\ \hline
            ABH only & $-1545.9\pm0.2$ & 0\\ 
		PBH+ABH - $\sigma=0.6$ & $-1515.4\pm0.2$ & $30.5\pm0.3$ \\ 
            \hline
            \multicolumn{3}{c}{All events except GW200210\_092254 and $\mmin=5~\msun$} \\ \hline
            ABH only & $-1516.8\pm0.2$ & 0\\ 
		PBH+ABH - $\sigma=0.6$ & $-1515.2\pm0.2$ & $1.66\pm0.3$ \\ 
            \hline\hline
	\end{tabular}
      \caption{Evidence and Bayes factors for the different fits carried out.}
      \label{tab:Bayesfactors}
\end{table}

\begin{figure*}
    \centering
    \includegraphics[width=0.92\textwidth]{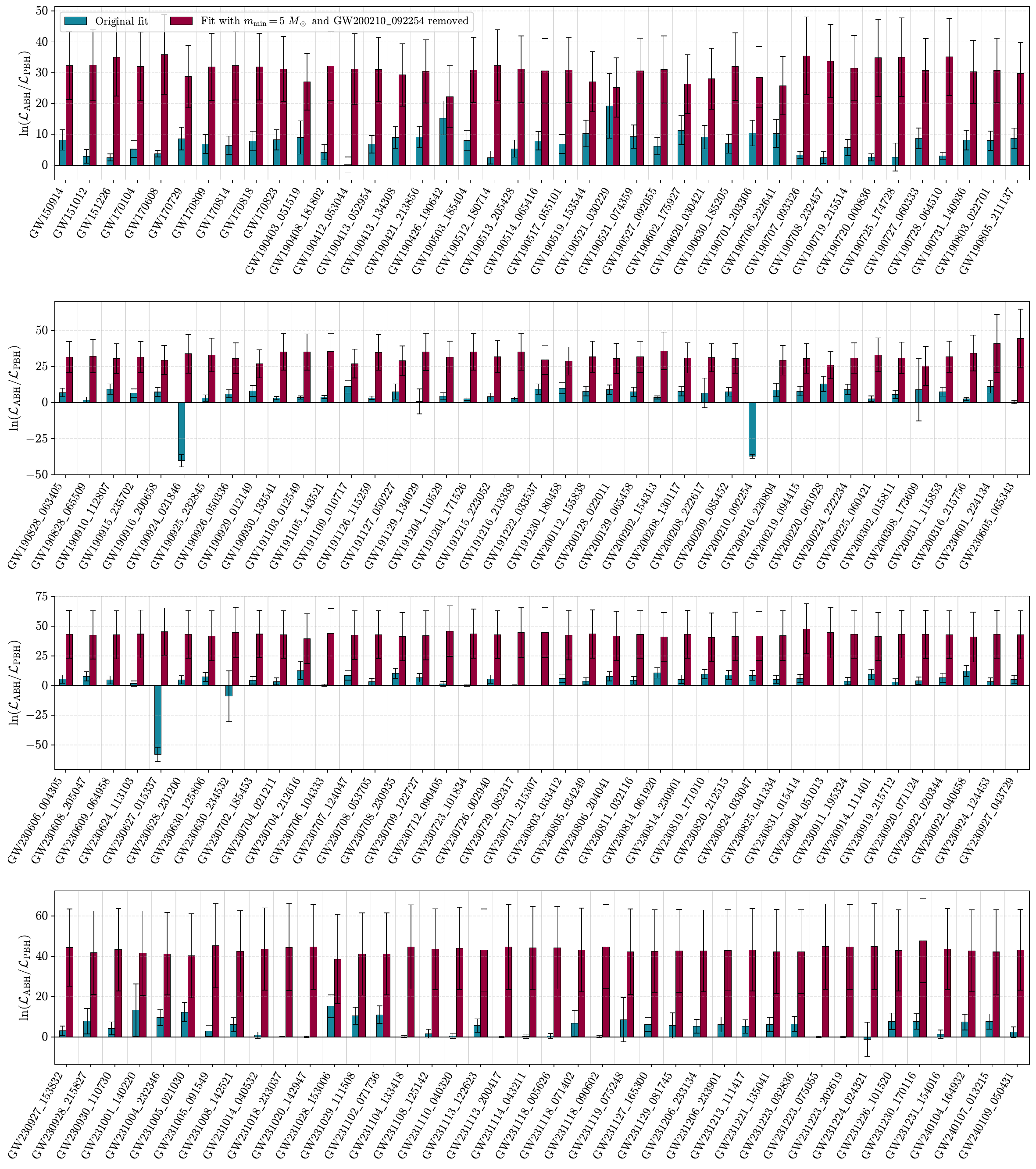}
    \caption{Log–likelihood ratio of the ABH and PBH event likelihoods, for the events considered with the ABH+PBH fit with a PL+2P as the ABH parametrisation and the log-normal with $\sigma=0.6$ for the PBH one. Positive values show that the event fits well with the ABH parametrisation while negative values indicate, that it is more likely to belong to the PBH binary population. The colour represents the two first considered: the one with the full set of events and leaving $\mmin$ as a free parameter and the fit where GW200210\_092254 is excluded and $\mmin = 5~\msun$. }
    \label{fig:PBHvsABH_sig06}
\end{figure*}

Regarding PBHs, since the contribution from the early three-body channel is dominant only at very high or low masses, we include it only in Fig.~\ref{fig:LimitsO4a_Combined}. For the combined ABH+PBH fit, which is restricted to the solar-mass range where LVK has detected events, we employ only the two-body channel, as it is several orders of magnitude stronger than the three-body channel at these masses. For the PBH mass function, we consider the log-normal distributions characterised by the peak masses (modes) $m_c$ and widths $\sigma$. The PBH population is modelled with three parameters,
\be
    \Lambda_{\rm PBH} = \{\fpbh, m_c, \sigma \} \,.
\ee

The priors used for the combined fit of primordial and astrophysical populations are given in Table~\ref{tab:priors}. To sample the posterior, we use the nested sampler \texttt{dynesty}~\cite{Speagle:2019ivv} and \texttt{bilby}~\cite{Ashton:2018jfp} as the high-level wrapper.

\begin{figure*}
    \centering
    \includegraphics[width=\textwidth]{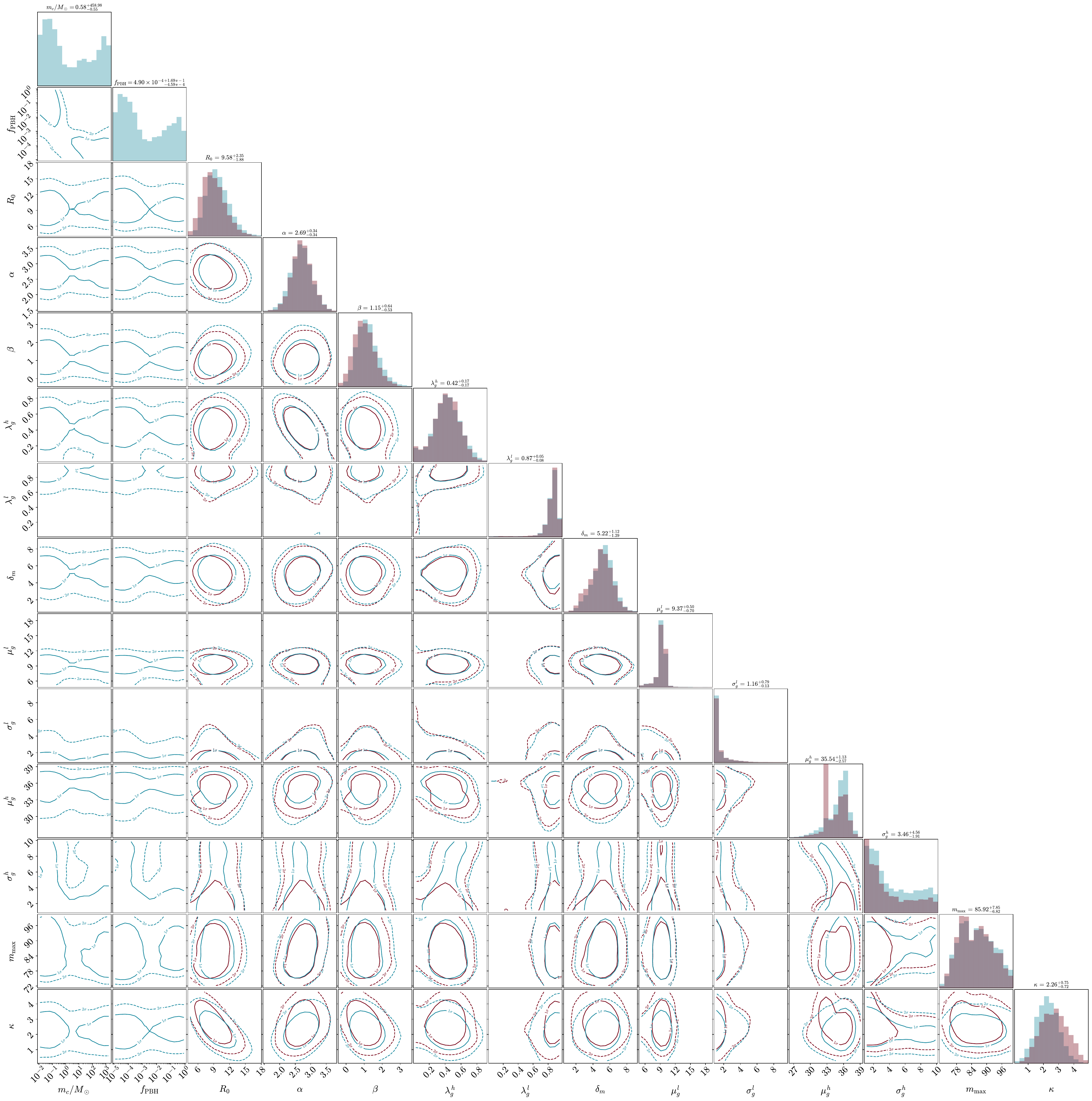}
    \caption{Posteriors for the BH binary merger rate model combining PBH and ABH binaries (blue) and for the model containing only ABH binaries (red) when $m_{\min}=5~M_\odot$ and excluding the event GW200210\_092254. The PBH binaries are assumed to have a log-normal mass function with width $\sigma = 0.6$ and the ABH binaries are described by a phenomenological PL+2P model given in appendix~\ref{app:parametrizedABHs}. $R_0$ in Gpc$^{-3}$yr$^{-1}$.}
    \label{fig:CornerPlot_mmin5_excl_events}
\end{figure*}

The first fit that we perform uses only the ABH parametrisation with all the parameters free, as specified in Table~\ref{tab:priors}, with all the available events up to O4a and a false alarm rate smaller than 1/year to establish the baseline case for comparison. Then, we add the PBH population with $\sigma = 0.6$ and repeat the fit. Figure~\ref{fig:CornerPlot} shows the full corner plot with the results of these two fits, and the Bayesian evidences are reported in Table~\ref{tab:Bayesfactors}. The main difference observed is that the astrophysical population is very different when considered alone compared to when the PBH is added on top. The biggest change is observed in the $\mmin$ parameter. For the ABH-only fit, the posterior tends towards the lower limit of the prior support, while in the case where both ABH and PBH populations are considered, its value shifts towards the upper limit. The position and width of the Gaussian peaks are also significantly altered.

Regarding the PBH population, the variables converge towards a very small region of the parameter space, clustered around $\langle m \rangle = 8.06~\msun$ and $\fpbh = 7.3\times 10^{-4}$. This is not the behaviour that was observed in the similar analysis of Ref.~\cite{Andres-Carcasona:2024wqk}, where only events up to O3b were considered. In this case, the fit clearly prefers to have a subdominant population of PBHs on top of the ABHs, which helps fitting the data better. The log-Bayes factor of $\ln \mathcal{B}_{\rm ABH}^{\rm PBH+ABH} = 30.4\pm 0.3$ confirms this. The main reason for this result lies in the mass distribution of the events. Since the PBH mass function is flexible in the sense that it can possess a peak at an arbitrary mass, it can accommodate some of the features of the mass spectrum. In this particular case, the fit is trying to explain the lowest mass event using the PBH mass function and employs the ABH one to explain the rest. Since the model that we are using for the ABH population is phenomenological and has several parameters that add flexibility to it, the data prefers to use each population to explain a different region of the mass parameter space. 

To support this, we can look at the individual likelihoods per event. Each individual likelihood is defined as shown in Eq.~\eqref{eq:individual_likelihood} but we can use either the PBH or the ABH merger rate model evaluated at each posterior sample. We take $100$ posterior samples to estimate the error in the quoted value as well. The log-likelihood ratio between ABH and PBH is displayed in Fig.~\ref{fig:PBHvsABH_sig06} in blue.  The events that are picked up as primordial--mainly GW190924\_021846, GW200210\_092254, GW230627\_015337, GW230630\_234532-- have one or both masses towards the low end of the parameter space, with the extreme case of GW200210\_092254, which contains the secondary mass in the lower mass gap. Therefore, the sampler is fitting these low-mass events with the PBH log-normal mass function, and the rest, which have higher masses (and thus the parameter $\mmin$ shifts to higher values), are accounted for using the ABH parametrisation.

Since we note that this is a caveat of using a phenomenological model for the ABH instead of a more physics-informed one, we now fix $\mmin=5~\msun$ which is supported by the current astrophysical knowledge and exclude GW200210\_092254 from our analysis to avoid dealing with a lower mass gap event. We show the full corner plot in Fig.~\ref{fig:CornerPlot_mmin5_excl_events}, which presents a much more reasonable result. In this case, the ABH parameters are compatible for both fits, and the PBH population exhibits an exclusion region, as shown in Fig.~\ref{fig:withABH}.

\begin{figure}
\vspace{-12pt}
    \includegraphics[width=0.49\columnwidth]{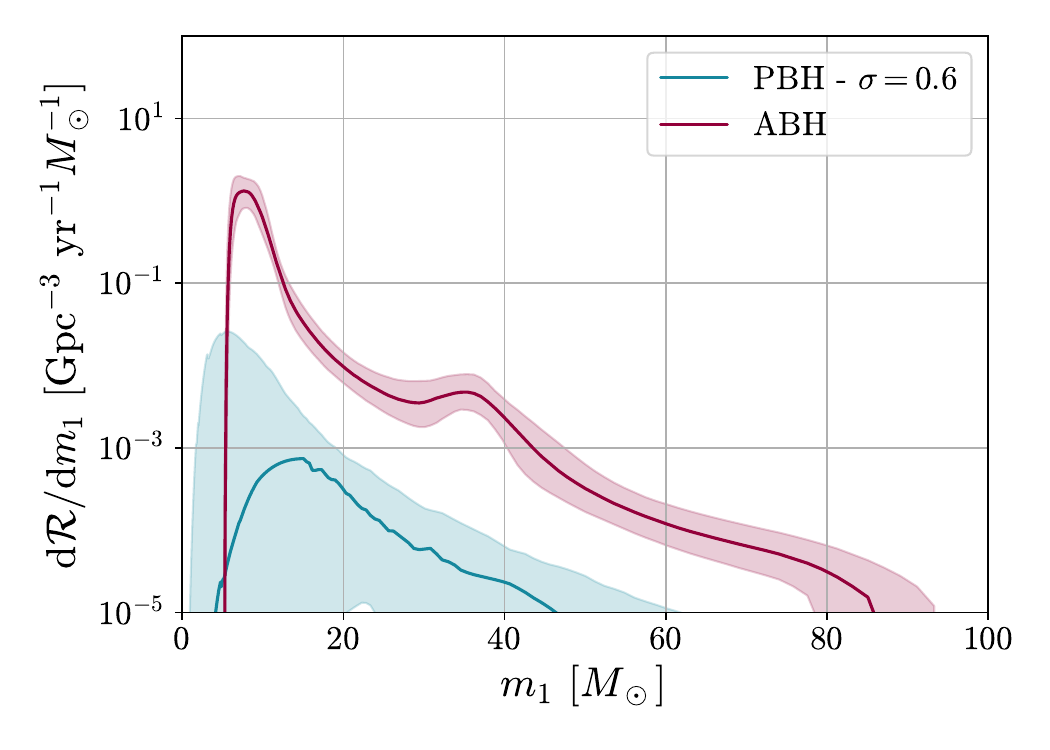} 
    \includegraphics[width=0.49\columnwidth]{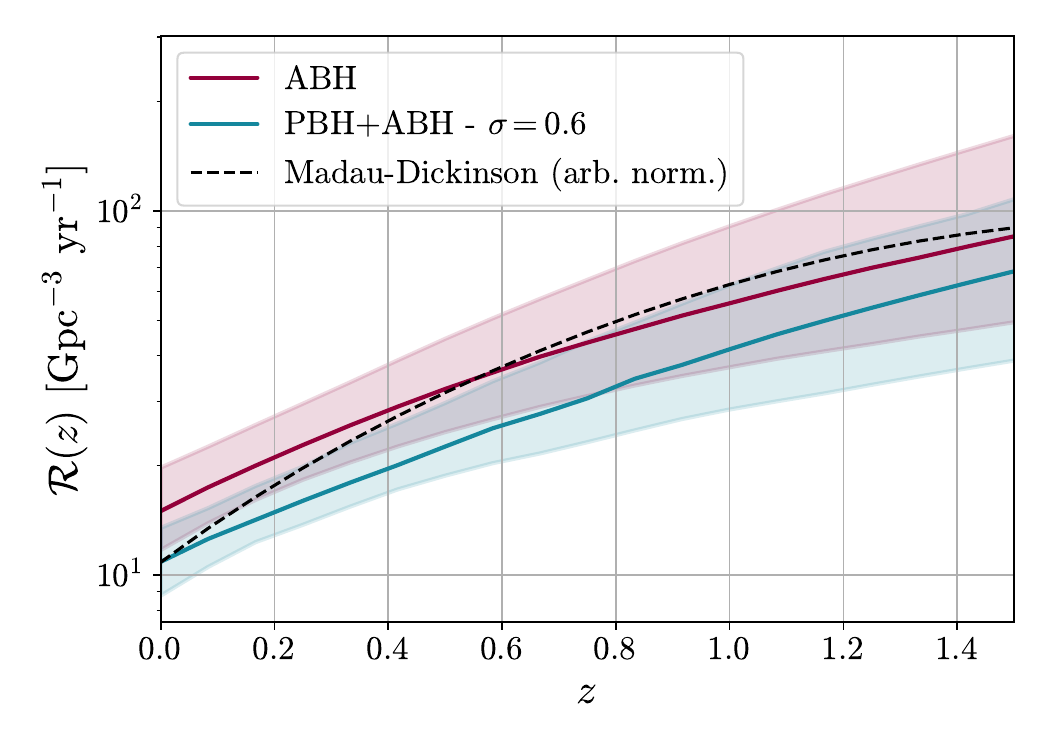}
    \vspace{-6pt}
    \caption{The merger rate as a function of the primary mass at $z=0$ (upper panel) and redshift (lower panel) for the ABH population in the PL+2P parametrisation (red) and for the PBH population with $\sigma=0.6$ (blue). The minimal ABH mass is fixed to $m_{\min}=5~M_\odot$ and the event GW200210\_092254 is omitted.}
    \label{fig:Rz_dRdm1_combined}
    \vspace{-8pt}
\end{figure}

The upper panel of Fig.~\ref{fig:Rz_dRdm1_combined} shows the differential merger rate as a function of the primary mass at $z=0$ and the redshift evolution for the ABH-only model and the PBH+ABH model with $\sigma=0.6$. The ABH population shows the expected behaviour of the smooth power law plus a peak at $\sim 35~\msun$, with an excess around $\sim 10~\msun$, thanks to the increased statistics of the O4a run~\cite{LIGOScientific:2025pvj}. The contribution of PBHs to the merger rate is several orders of magnitude smaller across the relevant mass range, reflecting the upper bound obtained on the PBH abundance. From the evolution of the merger rate with redshift, shown in the lower panel of Fig.~\ref{fig:Rz_dRdm1_combined}, we see that the merger rate for both ABHs and PBHs in this redshift range exhibits an evolution consistent with the star formation rate~\cite{Madau:2014bja}.

\begin{figure*}
    \centering
    \includegraphics[width=0.49\textwidth]{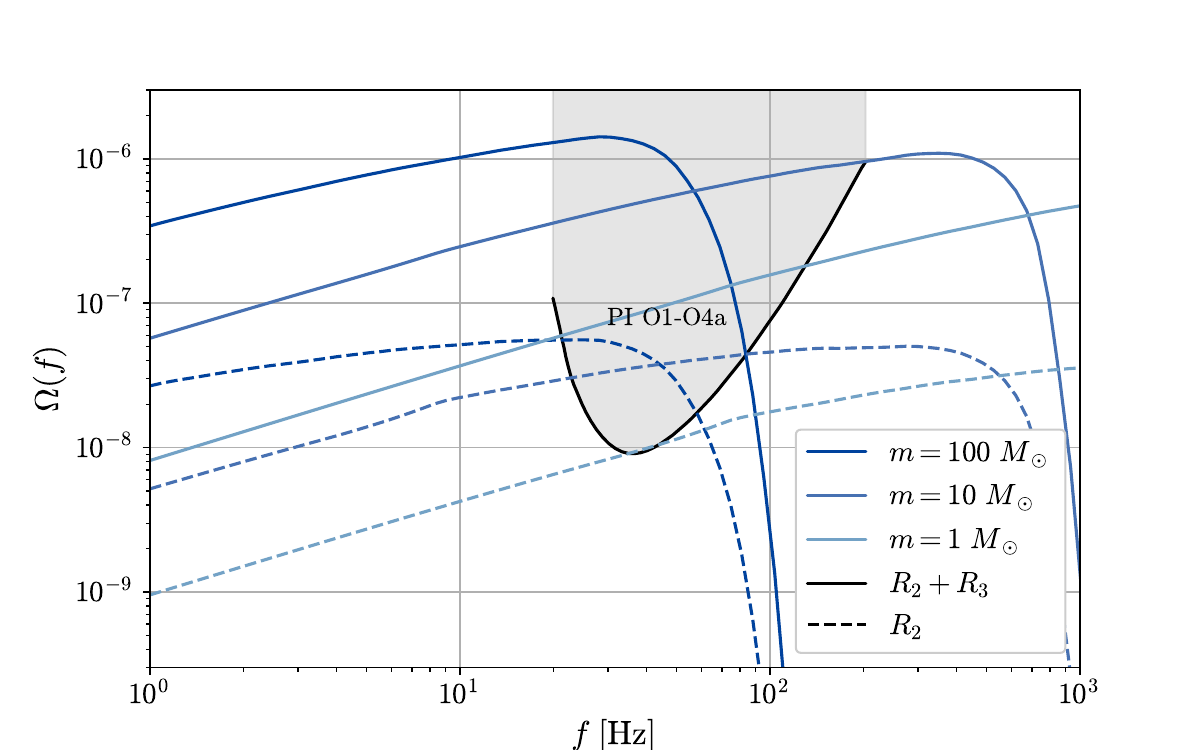}
    \includegraphics[width=0.49\textwidth]{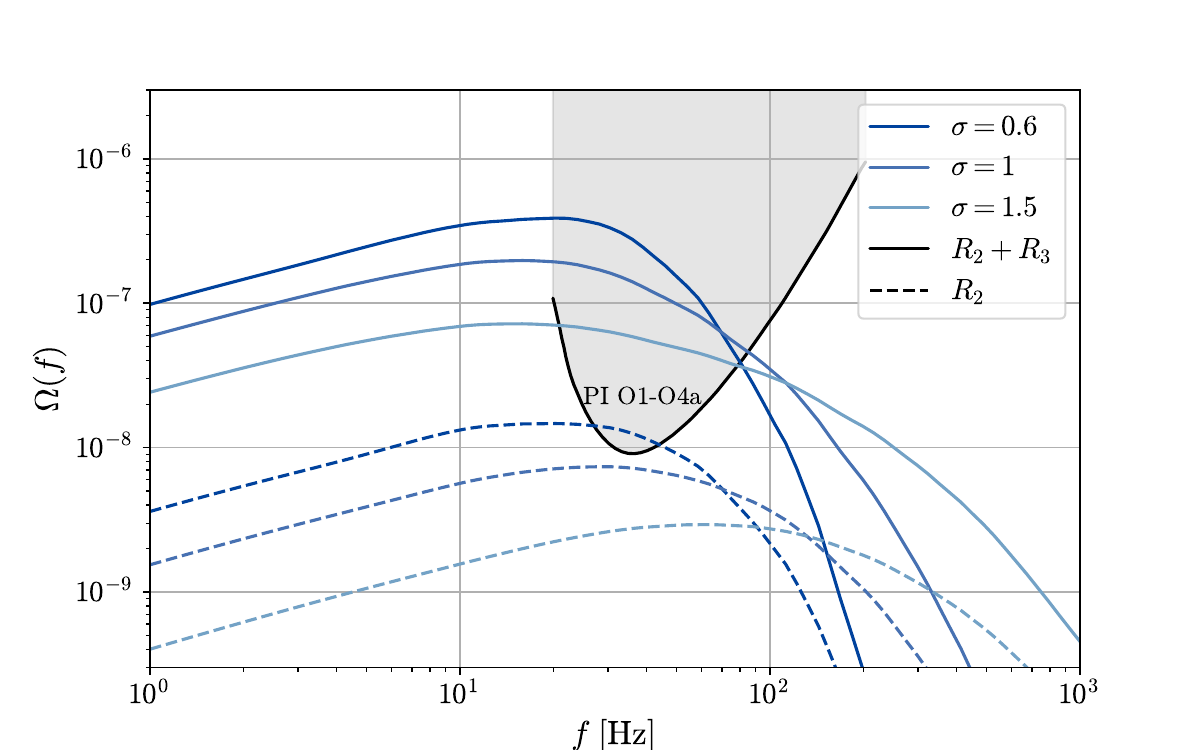}
    \caption{(\textit{left}) SGWB spectrum for the monochromatic mass function for different peak masses including the two- and three-body mechanisms for $f_{\rm PBH} = 1$. (\textit{right}) SGWB spectrum for the log-normal mass function for different widths including the two- and three-body mechanisms for $f_{\rm PBH} =1$ and $m_c = 100~M_\odot$. The black curve indicates the $2\sigma$ CL power-law integrated LVK sensitivity~\cite{LIGOScientific:2025bgj}.}
    \label{fig:SGWB_Omega}
\end{figure*}

\section{Modelling the stochastic GW background from binaries}
\label{app:SGWB}

Evaluation of the SGWB spectrum requires estimation of the inspiral-merger-ringdown GW energy spectrum. We implement a parametrisation used for the \texttt{IMRPhenom} family of waveforms, which reads as~\cite{Ajith:2009bn}
\bea \label{eq:dEdF_IMR_piecewise}
  &\frac{\td E}{\td f_s} = \frac{\pi^{2/3}}{3} \mathcal{M}_c^{5/3} f_s^{-1/3}
  \begin{cases}
    \big(1+\alpha_2u^2\big)^2, & f_s < f_1,\\
    w_1 f_s\,\left(1+\varepsilon_1 u+\varepsilon_2 u^2\right)^2, & f_1 \le f_s < f_2,\\
    w_2 f_s^{7/3}\dfrac{f_4^{\,4}}{\left[4(f_s-f_2)^2+f_4^{2}\right]^2}, & f_2 \le f_s < f_3,
  \end{cases}
\eea
and is valid for a quasi-circular, non-spinning binary. Here, the variable $u$ is defined as $u = (\pi M f_s)^{1/3}$, where $M = m_1 + m_2$, and the break frequencies $f_i$, with $i=1,2,3,4$, as $f_i = u_i^3/\pi M$ with
\bea \label{eq:ui_poly}
    u_1^{3} &= 0.066 + 0.6437\eta - 0.05822\eta^2 - 7.092\eta^3 \,, \\
    u_2^{3} &= 0.185 + 0.1469\eta - 0.0249\eta^2 + 2.325\eta^3 \,, \\
    u_3^{3} &= 0.3236 - 0.1331\eta - 0.2714\eta^2 + 4.922\eta^3 \,, \\
    u_4^{3} &= 0.0925 - 0.4098\eta + 1.829\eta^2 - 2.87\eta^3 \,,
\eea
where $\eta$ denotes the symmetric mass ratio of the binary. Finally, the inspiral correction and merger-shape coefficients are~\cite{Ajith:2009bn}
\be \label{eq:alpha_eps}
    \alpha_2 = -\frac{323}{224} + \frac{451}{168}\eta \,, \quad
    \varepsilon_1 = -1.8897 \,, \quad
    \varepsilon_2 = 1.6557 \,,
\ee
while the factors $w_1$ and $w_2$ enforce continuity (and overall normalization) across the transition frequencies by being set to
\be \label{eq:w12}
  w_1 = f_1^{-1}
  \frac{\big[1+\alpha_2 u_1^{2}\big]^2}{\left[1+\varepsilon_1 u_1+\varepsilon_2 u_1^{\,2}\right]^2}, \qquad
  w_2 = w_1 f_2^{-4/3}\left[1+\varepsilon_1 u_2+\varepsilon_2 u_2^{2}\right]^2.
\ee

The computation of the SGWB includes a three-dimensional integral over the masses and redshift. This can be computationally expensive and, therefore, we integrate it using the approach described in Ref.~\cite{Turbang:2023tjk}. We start by drawing $N_s$ samples from a uniform distribution for the parameters $z$, $m_1$ and $m_2$. Each $i^{\rm th}$ parameter sample has its corresponding drawing probability of $\pi_{\mathrm{draw}}(z^i,m_1^i,m_2^i)$. For each sample, we compute its energy spectrum ${\td E^i/\td f_s}$ following Eq.~\eqref{eq:dEdF_IMR_piecewise}. We can then use these samples to estimate Eq.~\eqref{eq:Omega_GW} via a Monte Carlo integration, but we need to correct the difference in the populations used to obtain these samples and the population described by the hyperparameters $\Lambda$. We reweigh the different samples of the individual energy spectrum of the binaries and evaluate the SGWB spectrum as
\be
    \Omega(f) \approx \frac{1}{N_s}\sum_{i=1}^{N_{\rm non~det}}w_i \frac{\td E^i}{\td f_s}\, .
    \label{eq:Omega_MC}
\ee
Notice how the sum is done only for the $N_{\rm non~det}$ events that are not detected, to account for the selection effects imposed by the $1-p_{\rm det}$ term. The weights correspond to 
\be
w_i = \frac{\mathrm{d}V_c^i}{\mathrm{d}z}\frac{\mathrm{d}R^i}{\mathrm{d}m_1\mathrm{d}m_2}\frac{1}{4\pi D_L(z^i)^2}\frac{1}{\pi_{\mathrm{draw}}(z^i,m_1^i,m_2^i)}\, .
\ee

We evaluate Eq.~\eqref{eq:Omega_MC} for the monochromatic and log-normal mass functions to see the SGWB spectrum recovered. We show in Fig.~\ref{fig:SGWB_Omega} both cases for different parameters of the mass functions. Additionally, we compare the expected SGWB when both the three- and two-body mechanisms are included and when only the two-body one is considered. The addition of the three-body mechanism enhances the expected background, as it adds many contributions from unresolvable binaries in the low-mass regime.

\begin{figure}
    \centering
    \includegraphics[width=0.6\columnwidth]{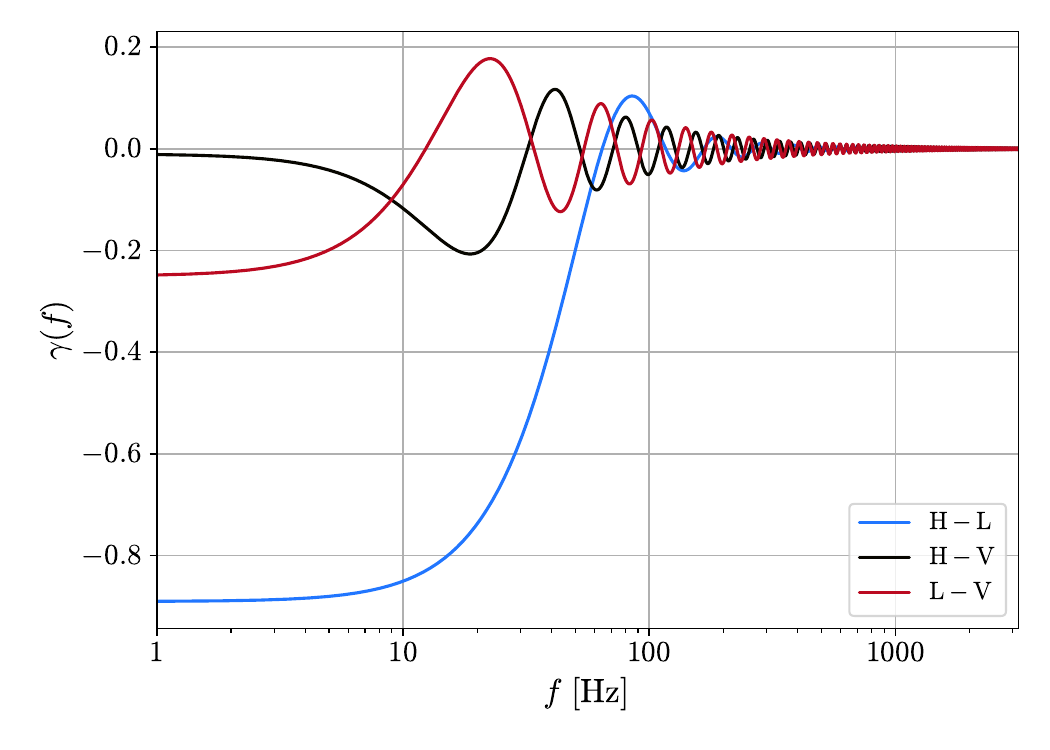}
    \caption{Overlap reduction function used for the evaluation of the SNR of the SGWB.}
    \label{fig:ORF}
\end{figure}

Finally, to evaluate the SNR of Eq.~\eqref{eq:SNR_cc_SGWB} the overlap reduction function (ORF) is needed. The ORF captures both the reduction in sensitivity due to the physical separation of detectors $I$ and $J$ and the relative orientation of their arms. The response of a pair of detectors to an isotropic, unpolarised stochastic background is encoded in this ORF as~\cite{Christensen:1992wi,Allen:1997ad}
\be \label{eq:ORF_def}
  \gamma_{IJ}(f)
  = \frac{5}{8\pi} \sum_A \int_{S^2} \td \hat{\mathbf{n}}
  e^{2\pi i f\, \hat{\mathbf{n}}\cdot \Delta \mathbf{x}_{IJ}/c}\,
  F_I^A(\hat{\mathbf{n}}) F_J^A(\hat{\mathbf{n}}),
\ee
where the integration is taken over all sky directions $\hat{\mathbf{n}}$, and the sum is carried over the GW polarizations $A \in \{+,\times\}$. The other terms are $\Delta \mathbf{x}_{IJ} = \mathbf{x}_I - \mathbf{x}_J$, which is the separation vector between the detector locations, $\hat{\mathbf{n}}$, the pointing vector of the incoming GW and $F_I^A(\hat{\mathbf{n}})$ and $F_J^A(\hat{\mathbf{n}})$, the antenna pattern functions of detectors $I$ and $J$ for polarization $A$. The prefactor $5/8\pi$ ensures that $\gamma_{IJ}(0)=1$ for a pair of co-located, co-aligned detectors~\cite{Romero-Rodriguez:2024ldc}. We show in Fig.~\ref{fig:ORF} the ORF for the two LIGO detectors and Virgo.

\bibliographystyle{JHEP}
\bibliography{ref}

\end{document}